\documentclass{article}
\usepackage{amssymb,amsmath}
\usepackage{amsfonts}
\usepackage[latin1]{inputenc}
\usepackage{fancyhdr}
\usepackage{yfonts}
\usepackage{mathrsfs}
\usepackage[dvips]{graphicx}
\usepackage[rflt]{floatflt}
\usepackage{amsthm}
\usepackage{cite}

\newtheorem{prop}{Proposition}
\newtheorem{thm}{Theorem}

\newtheorem{cor}{Corollary}

\newtheorem{lem}{Lemma}
\newtheorem{rem}{Remark}

\providecommand{\prt}[1]{\left( #1 \right)}
\providecommand{\abs[1]}{\left|#1\right|}

\providecommand{\quotes[1]}{``#1"}

\providecommand{\T}{\mathcal{T}}%
\renewcommand{\S}{\mathcal{S}}%
\providecommand{\A}{\mathcal{A}}%
\providecommand{\X}{\mathcal{X}}%
\providecommand{\Y}{\mathcal{Y}}%

\title{Primitive operations for the construction and reorganization of minimally persistent formations}

\author{Julien M. Hendrickx\thanks{J.\ M.\ Hendrickx (corresponding author) and V.\ Blondel are
with Department of Mathematical Engineering, Universit\'e
catholique de Louvain, Avenue Georges Lemaitre 4, B-1348
Louvain-la-Neuve, Belgium; {\tt\small
hendrickx,blondel@inma.ucl.ac.be}. Their work is supported by the
Belgian Programme on Interuniversity Attraction Poles initiated by
the Belgian Federal Science Policy Office, and The Concerted
Research Action (ARC) \quotes{Large Graphs and Networks} of the
French Community of Belgium. The scientific responsibility rests
with its authors. Julien Hendrickx holds a FNRS fellowship
(Belgian Fund for Scientific Research). This work was supported in
part by DoD AFOSR URI for ''Architectures for Secure and Robust
Distributed Infrastructures,'' F49620-01-1-0365 (led by Stanford
University).}, Bar\i\c{s} Fidan\thanks{B.\ Fidan, C.\ Yu and B.\
Anderson are with Australian National University and National ICT
Australia , 216 Northbourne Ave, Canberra ACT 2601 Australia ;
{\tt\small baris.fidan,brad.yu,brian.anderson@nicta.com.au}. Their
work is supported by an Australian Research Council Discovery
Project Grant and by National ICT Australia, which is funded by
the Australian Government's Department of Communications,
Information Technology and the Arts and the Australian Research
Council through the Backing Australia's Ability Initiative.
Changbin Yu is an Australia-Asia Scholar supported by the
Australian Government's Department of Education, Science and
Training through Endeavours Programs.}, Changbin Yu,\\ Brian D.O.
Anderson and Vincent D. Blondel }

\begin{document}
\maketitle
\begin{abstract}
In this paper, we study the construction and transformation of
two-dimensional persistent graphs. Persistence is a generalization
to directed graphs of the undirected notion of rigidity. In the
context of moving autonomous agent formations, persistence
characterizes the efficacy of a directed structure of unilateral
distances constraints seeking to preserve a formation shape.
Analogously to the powerful results about Henneberg sequences in
minimal rigidity theory, we propose different types of directed
graph operations allowing one to sequentially build any minimally
persistent graph (i.e. persistent graph with a minimal number of
edges for a given number of vertices), each intermediate graph
being also minimally persistent. We also consider the more generic
problem of obtaining one minimally persistent graph from another,
which corresponds to the on-line reorganization of an autonomous
agent formation. We prove that we can obtain any minimally
persistent formation from any other one by a sequence of
elementary local operations such that minimal persistence is
preserved throughout the reorganization process.
\end{abstract}

\section{Introduction} \label{sec:intro}

The recent progress in the field of autonomous agent systems has
led to new problems in control theory
\cite{BaillieulSuri:2003,DasSpletzerKumarTaylor:2002} and graph
theory
\cite{ErenWhiteleyAndersonMorseBelhumeur:2005,OlfatiMurray:2002,HendrickxAndersonDelvenneBlondel:2005}.
By autonomous agent, we mean here any human controlled or unmanned
vehicle that can move by itself and has a local intelligence or
computing capacity, such as ground robots, air vehicles or
underwater vehicles. The results derived in this paper concern
mostly autonomous agents
evolving in a two dimensional space.\\%

\begin{figure}
\centering
\begin{tabular}{cc}
\includegraphics[scale = .25]{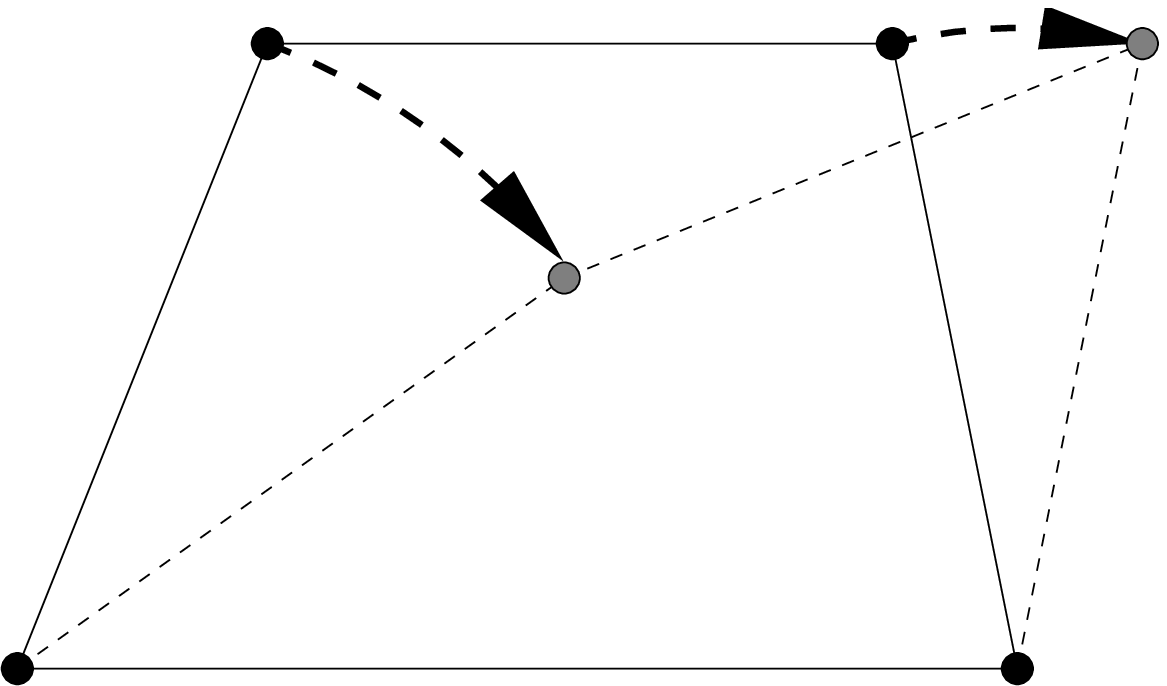}&
\includegraphics[scale = .25]{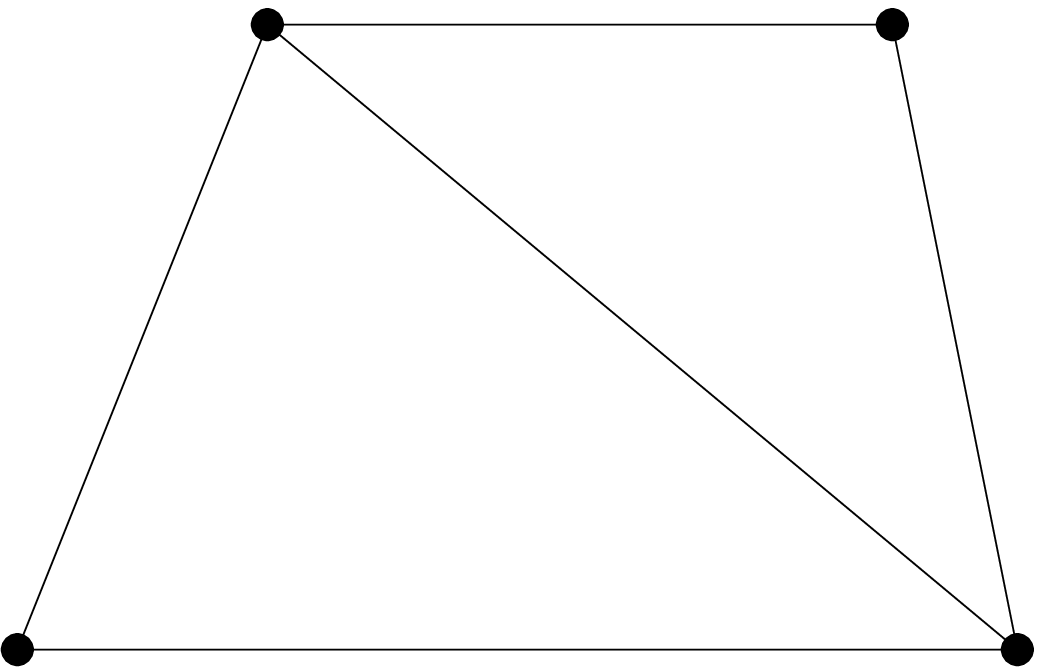}
\\(a)&(b)
\end{tabular}
\caption{Representation of (a) a non-rigid and (b) a rigid
graph/formation. The solid structure in (a) can indeed be deformed
to the dotted structure without breaking any distance constraint.}
\label{fig:rigidity}
\end{figure}

\begin{figure}
\centering
\begin{tabular}{cc}
\includegraphics[scale = .37]{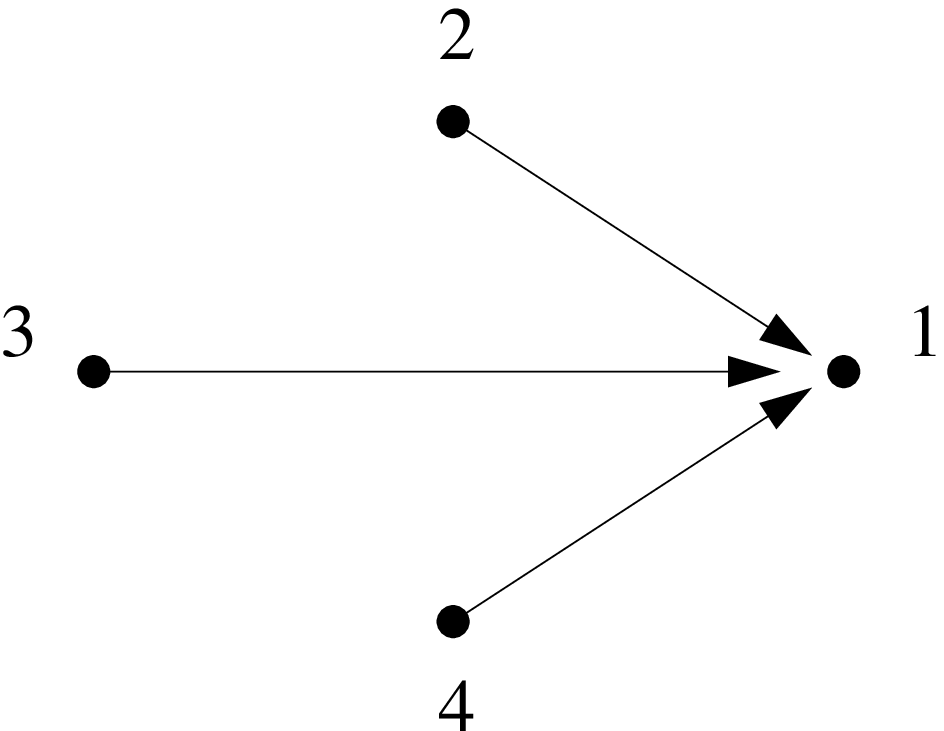}&
\includegraphics[scale = .37]{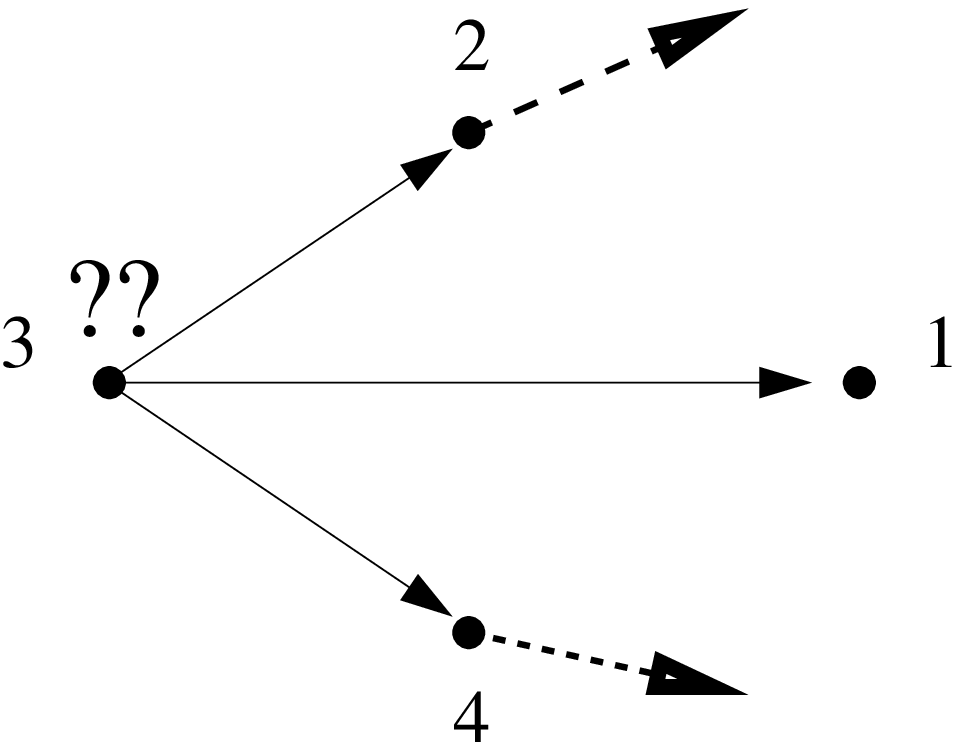}
\\(a)&(b)
\end{tabular}
\caption{Representation of (a) a constraint consistent and (b) a
non-constraint consistent (in 2 dimension) graph/formation. One
can indeed see in (b) that for almost any uncoordinated continuous
displacement of the agents 2 and 4 (which are unconstrained), the
agent 3 is unable to move in such a way that it maintains its
distances to all of 1,2 and 4 constant. However, such a situation
could not happen in graph (a).} \label{fig:constr_consist}
\end{figure}

\begin{figure}
\centering
\includegraphics[scale = .4]{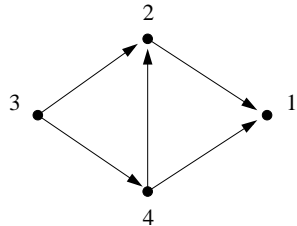}
\caption{Representation of a persistent graph, i.e., a rigid
constraint consistent graph.} \label{fig:persi}
\end{figure}

Many applications require some inter-agent distances to be kept
constant during a continuous move in order to preserve the shape
of a multi-agent formation. In other words, some inter-agent
distances are explicitly maintained constant so that all the
inter-agent distances remain constant. The information structure
arising from such a system can be efficiently modelled by a graph,
where agents are abstracted by vertices and actively constrained
inter-agent distances by edges. We assume here that those
constraints are unilateral, i.e., that the responsibility for
maintaining a distance is not shared by the two concerned agents
but relies on only one of them while the other one is unaware of
it. This asymmetry is modelled by the use of directed edges in the
graph. The characterization of the directed information structures
which can efficiently maintain the formation shape has begun to be
studied under the name of \quotes{directed rigidity}
\cite{BaillieulSuri:2003,ErenWhiteleyAndersonMorseBelhumeur:2005}.
These works included several conjectures about minimal directed
rigidity, i.e., directed rigidity with a minimal number of edges
for a fixed number of vertices. In
\cite{HendrickxAndersonDelvenneBlondel:2005}, Hendrickx et al.
proposed a theoretical framework to analyze these issues, where
the name of \quotes{persistence} was advanced in preference to
\quotes{directed rigidity}, since the latter notion does not
correspond to the immediate transposition of the undirected notion
of rigidity to directed graph. The intuitive definition of
persistence is the following: An information structure is
persistent if, provided that each agent is trying to satisfy all
the distance constraints for which it is responsible, all the
inter-agent distances remain constant and as a result the
formation shape is preserved. It is shown in
\cite{HendrickxAndersonDelvenneBlondel:2005} that persistence is
actually the conjunction of two distinct notions: rigidity of the
underlying undirected graph (i.e. the undirected graph obtained by
ignoring the direction of the edges), and constraint consistence.
Intuitively, rigidity means that, provided that all the prescribed
distance constraints are satisfied during a continuous
displacement, all the inter-agent distances remain constant, as
shown in Figure \ref{fig:rigidity}. Constraint consistence of an
information structure means that, provided that each agent is
trying to satisfy all its distances constraints, all the agents
actually succeed in doing so. In other words, no agent has an
impossible task, as shown in the example in Figure
\ref{fig:constr_consist}. Observe that this last notion depends
strongly on the directed structure of the graph, while rigidity
only relies on its underlying undirected graph. An example of
persistent graph is provided in Figure \ref{fig:persi}. Note that
for agents evolving in a two-dimensional space, a purely
combinatorial criterion to decide persistence is provided in
\cite{HendrickxAndersonDelvenneBlondel:2005}.


In this paper, we focus on minimally persistent graphs, that are
persistent graphs having a minimal number of edges (for a given
number of vertices), and their connections with minimally rigid
graphs. More particularly \emph{we analyze different ways to
sequentially build minimally persistent graphs, analogously to the
Henneberg sequences for the minimally rigid graphs}
\cite{Henneberg:11,TayWhiteley:85}. It has indeed long been known
that every minimally rigid graph can be obtained from the complete
graph on two vertices by a sequence of two basic operations, as
detailed in Section \ref{sec:Henneberg}. The natural extension of
these operations to directed graphs
\cite{ErenWhiteleyAndersonMorseBelhumeur:2005} does \emph{not}
allow one to build all minimally persistent graphs, as remarked in
\cite{HendrickxAndersonDelvenneBlondel:2005} and reviewed later in
Section \ref{sec:Henneberg}. We consider here different possible
additional operations that would help to achieve this purpose. We
also consider the more generic problem of obtaining one persistent
graph from another. From an autonomous agent point of view, this
corresponds to an on-line reorganization of the agent formation.
The subsequent analysis leads us then to the definition of
different \quotes{distances} between persistent graphs (the
distance between two graphs being the smallest number of
operations needed to obtain one from the other). Note that
although the notion of persistence has been also defined in three
or higher dimensions
\cite{YuHendrickxFidanAndersonBlondel:2005,HendrickxFidanYuAndersonBlondel:2005_conf,YuHendrickxFidanAnderson:2005},
the present analysis only concerns two-dimensional persistence,
i.e., the persistence of graphs representing the information
structure of a formation evolving in a two-dimensional space.
Extension to the three dimensional case may be difficult; even for
undirected graphs, Henneberg sequence theory is effectively
incomplete.
\\%

In Section \ref{sec:Henneberg}, we review the main properties of
minimally rigid and minimally persistent graphs. We present the
two basic undirected operations - vertex addition and edge
splitting - involved in the Henneberg sequences, together with
their natural extension to directed graphs. We show that although
these directed operations preserve minimal persistence, they are
not sufficient to build all minimally persistent graphs. This
analysis is done by reasoning on reverse construction of
persistent graphs using reverse operations. In Section
\ref{sec:third_op}, we show how the goal of building all minimally
persistent graph can be reached by introducing a third local
directed operation - edge reversal. We see that, unlike when
building minimally rigid undirected graphs with Henneberg
sequences, the required number of operations is not uniquely
determined by the size of the graph. We show in Section
\ref{sec:four_non_local} that this drawback can be avoided by
using only directed operations equivalent to the vertex addition
and the edge splittings from an undirected point of view. However,
we prove that a set of such operations allowing one to build all
minimally persistent graphs always contain at least one
non-confined operation, i.e. an operation reversing the directions
of (possibly several) edges that are not affected by the
corresponding operation for undirected graphs. We provide then
such a set of four operations, and analyze the relations between
this set and the set of three operations treated in Section
\ref{sec:third_op}. Finally, this paper ends with the concluding
remarks of Section \ref{sec:conclusions}.

\section{Directed and undirected Henneberg sequences} \label{sec:Henneberg}

In this section, we recall some results about (minimal) rigidity
and (minimal) persistence. We also describe the Henneberg
sequences for undirected graphs and show why their obvious
adjustment to the directed case is not sufficient to build all
minimally persistent graphs.

\subsection{Minimally rigid graphs}\label{sec:min_rig_graph}

Note that in this section, all graphs are considered as
undirected, but in the rest of this paper, they are always assumed
to be directed. However, although all the definitions and results
of this section are given for undirected graphs, they can also be
applied to directed graphs. If $G$ is a directed graph, we call
the \emph{underlying undirected graph} of $G$ the undirected graph
obtained by ignoring the directions of the edges
of $G$.\\%

The rigidity of a graph has the following intuitive meaning:
Suppose that each vertex represents an agent in a formation, and
each edge represents an inter-agent distance constraint enforced
by an external observer. The graph is rigid if for almost every
such structure, the only possible continuous moves are those which
preserve every inter-agent distance. Note that this notion also
represents the rigidity of a framework where the vertices
correspond to joints and the edges to bars. For a more formal
definition, the reader is referred to
\cite{TayWhiteley:85,HendrickxAndersonDelvenneBlondel:2005}. In
$\Re^2$, there exists a combinatorial criterion to check if a
given graph is rigid {(Laman's theorem
\cite{Laman:70,Whiteley:96}). A \emph{minimally rigid} graph is a
rigid graph such that no edge can be removed without losing
rigidity. From Laman's Theorem, it
is possible to deduce the following criterion:}%
\begin{prop}\label{prop:Laman_min}
A graph $G=(V,E)$ ($\abs{V}>1$) is minimally rigid if and only if
$\abs{E}=2\abs{V}-3$ and for all $E''\subseteq E, E'' \not=
\varnothing$, there holds $\abs{E''}\leq 2 \abs{V(E'')}-3$.
\end{prop}

We say that a pair of unconnected vertices defines an
\emph{implicit edge} in a graph $G=(V,E)$ if their connection
would create a subgraph $G'=(V',E')$ with $\abs{E'}>2\abs{V'}-3$.
Intuitively, this means that the addition of such an edge would
not improve the rigidity of the graph, i.e., the constraint that
this edge would enforce is a linear combination of already present
constraints. One can easily prove that two unconnected vertices
define an implicit edge in a graph if and only if there is a
minimally rigid subgraph containing both of them. By extension, we
sometimes call an edge of a graph an \emph{explicit edge}. In a
(minimally) rigid graph, every pair of vertices is connected by
either an explicit or implicit edge. But, if one removes an
(explicit) edge in a minimally rigid graph, the corresponding pair
of vertices never defines an implicit edge
in the graph obtained.\\%

\begin{figure}
\centering
\begin{tabular}{c}
\includegraphics[scale=.25]{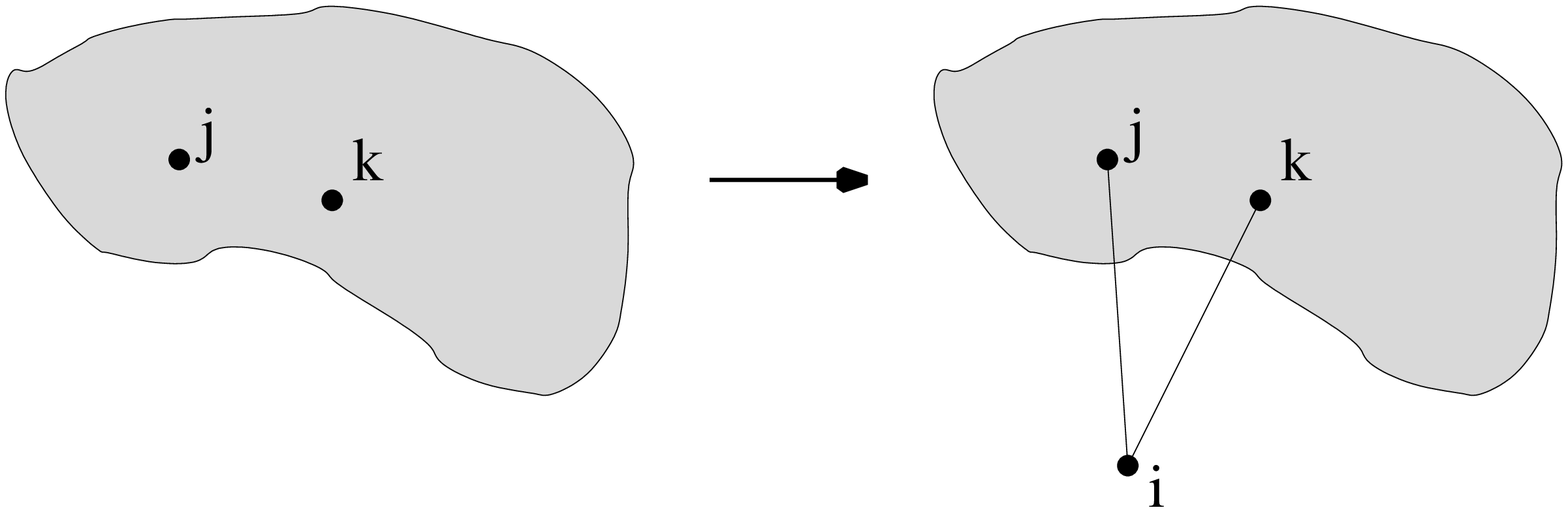}\\
(a)\\
\includegraphics[scale=.25]{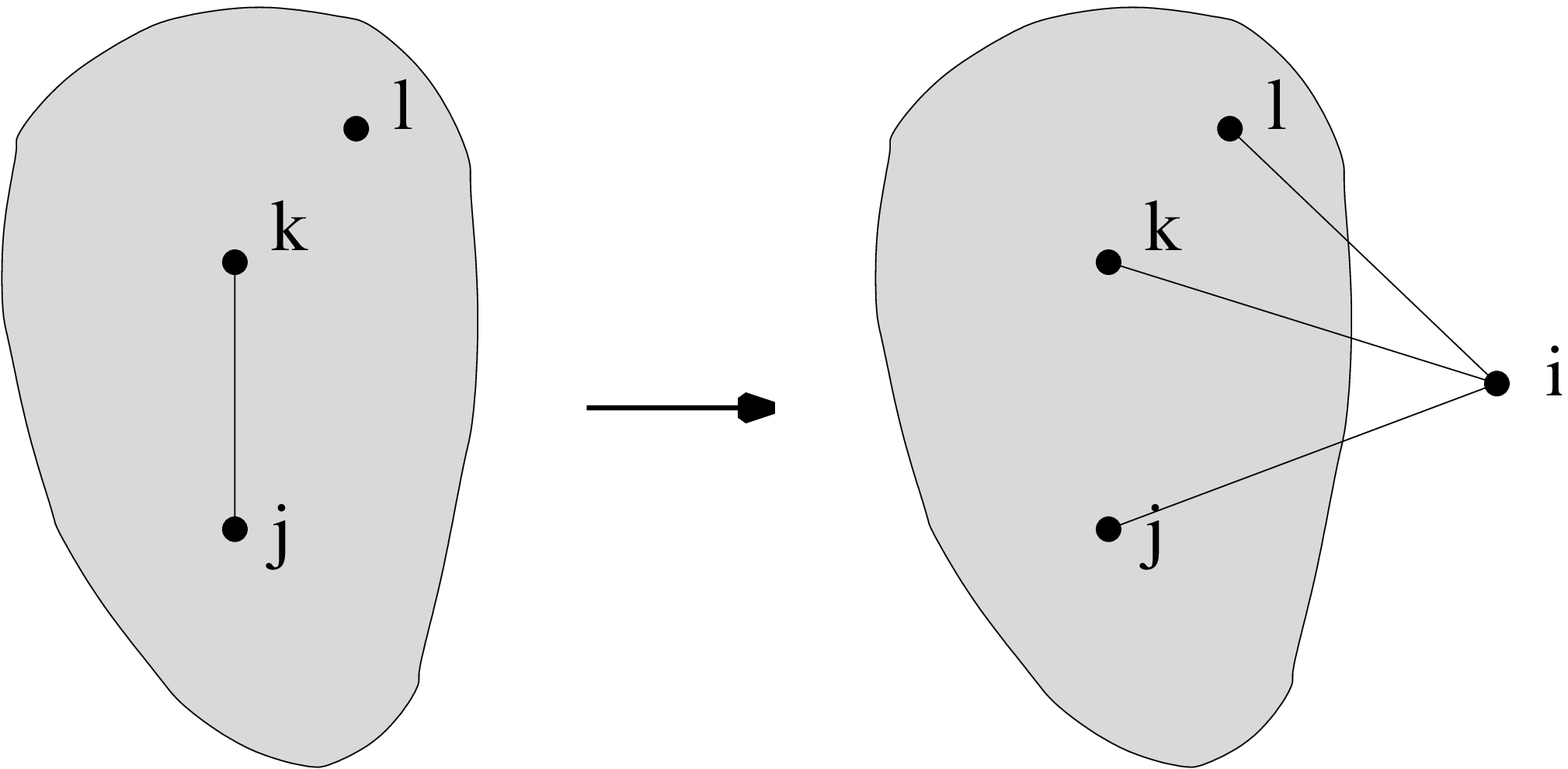}
\\
(b)\\
\end{tabular}
\caption{Representation of (a) undirected vertex addition
operation and (b) edge splitting
operation.}\label{fig:repres_undir_op}
\end{figure}

Let $j,k$ be two distinct vertices of a minimally rigid graph
$G=(V,E)$. A \emph{vertex addition} operation consists in adding a
vertex $i$, and connecting it to $j$ and $k$, as shown in Figure
\ref{fig:repres_undir_op}(a). One can see using Proposition
\ref{prop:Laman_min} that this operation preserves minimal
rigidity. Moreover, if a vertex $i$ has degree 2 in a minimally
rigid graph, one can always perform the inverse vertex addition
operation by removing it (and its incident edges) and obtain a
smaller minimally rigid graph.\\%

\begin{figure}
\centering
\begin{tabular}{ccc}
\includegraphics[scale=.3]{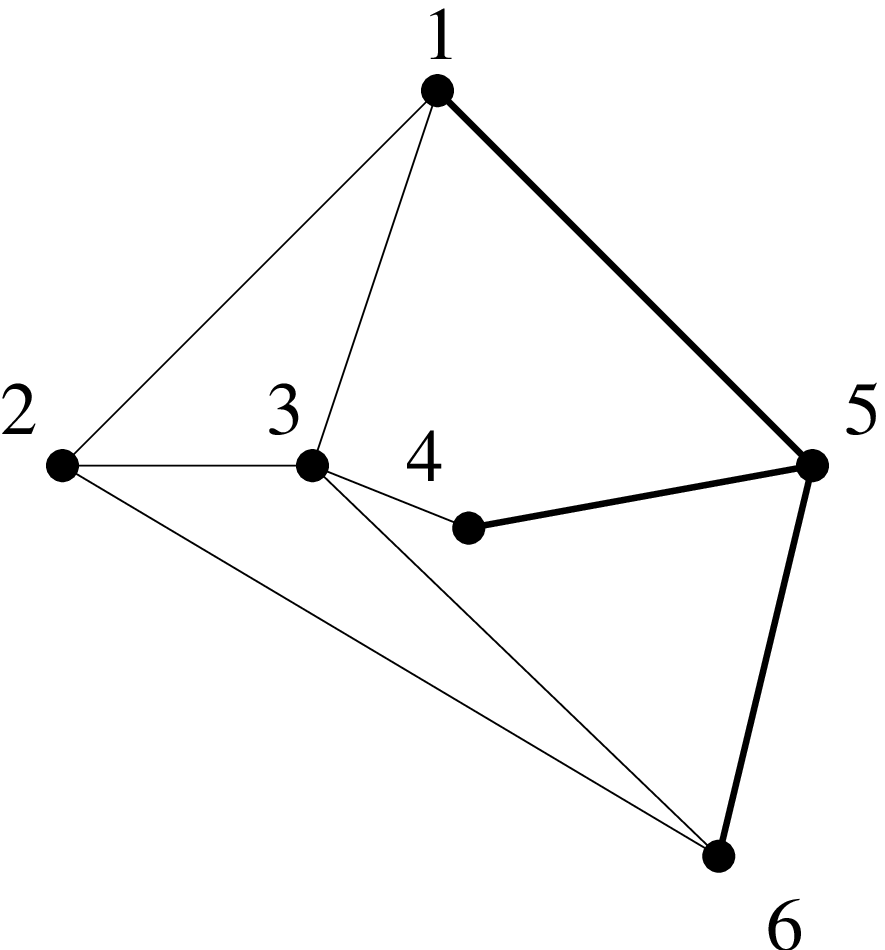}&
\includegraphics[scale=.3]{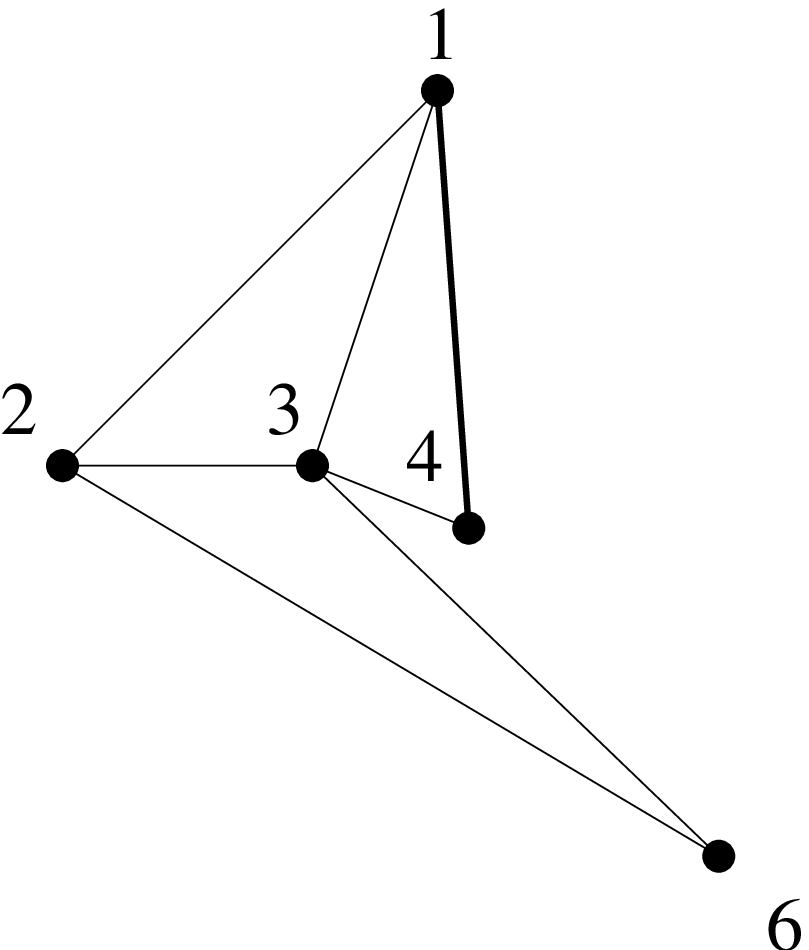}&
\includegraphics[scale=.3]{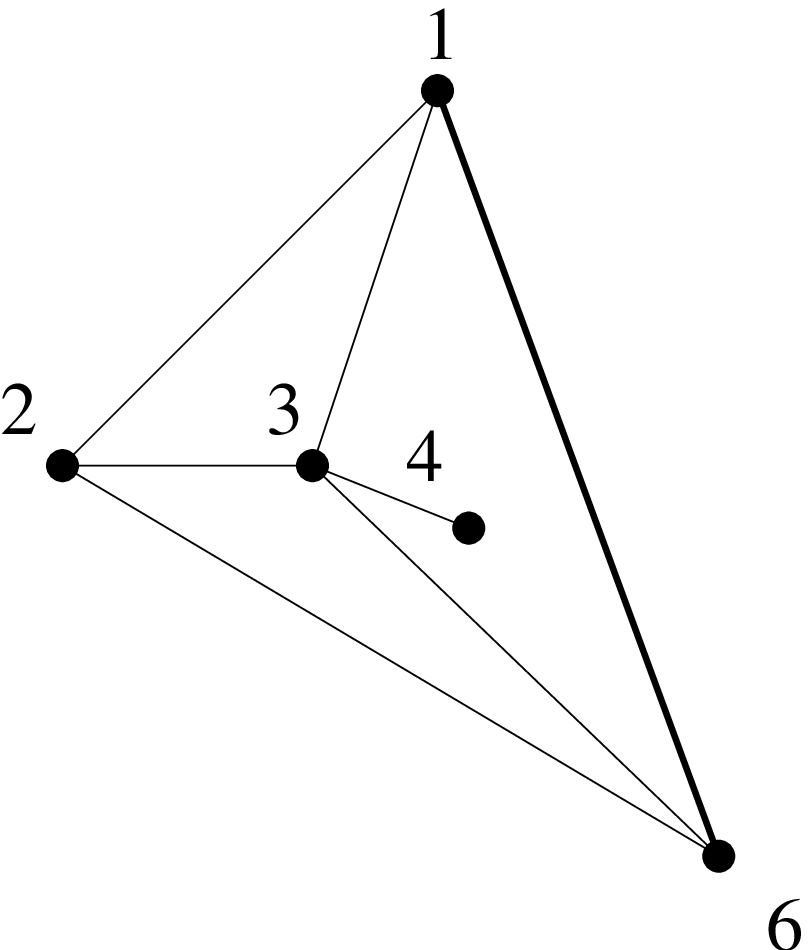}
\\(a)&(b)&(c)
\end{tabular}
\caption{Example of unfortunate added edge selection in reverse
edge splitting. After the removal of the vertex 5 from the
minimally rigid graph (a), minimal rigidity can be preserved by
the addition of the edge $(1,4)$ but not of $(1,6)$, as shown
respectively on (b) and (c). The pair $(1,6)$ defines an implicit
edge in the minimally rigid subgraph induced by 1, 2, 3 and 6.
}\label{fig:example_wrong_reverse_edgsplit}
\end{figure}

Let $j,k,l$ be three vertices of a minimally rigid graph such that
there is an edge between $j$ and $k$. An \emph{edge splitting}
operation consists in removing this edge, adding a vertex $i$ and
connecting it to $j$, $k$ and $l$, as shown in Figure
\ref{fig:repres_undir_op}(b). This operation provably preserves
minimal rigidity \cite{TayWhiteley:85}. The reverse operation is
less straightforward than the reverse vertex addition operation.
Given a vertex $i$ connected to $j$, $k$ and $l$, the minimal
rigidity of the graph is preserved if one removes $i$ and adds one
edge among $(j,k)$, $(k,l)$ and $(l,j)$. However, one cannot
always freely choose any one of these edges to add. One has indeed
to make sure that the added edge does not already belong to the
graph, and also that its addition does not create a subgraph
$G'=(V',E')$ with $\abs{E'}>2\abs{V'}-3$, i.e., that the pair of
vertices does not define an implicit edge in the graph obtained
after deletion of $i$. Figure
\ref{fig:example_wrong_reverse_edgsplit} shows an example of such
an unfortunate added edge selection. Suppose indeed that the
vertex 5 is removed from the minimally rigid graph
\ref{fig:example_wrong_reverse_edgsplit}(a). The pair $(1,4)$ does
provably not define an implicit edge, and its addition leads thus
to a minimally rigid graph, which is represented in Figure
\ref{fig:example_wrong_reverse_edgsplit}(b). However, if $(1,6)$
is added instead of $(1,4)$, the graph obtained contains a
subgraph $G'=(V',E')$ with $V'=\{1,2,3,6\}$ such that $6
=\abs{E'}>2\abs{V'}-3 = 5$, as shown in Figure
\ref{fig:example_wrong_reverse_edgsplit}(c). The pair $(1,6)$
defines thus an implicit edge. It is possible to prove that at
least one among the three possible pairs of vertices does not
define an actual nor an implicit edge
\cite{TayWhiteley:85,Laman:70}. One can thus always perform a
reverse edge splitting on any vertex with a degree 3.\\%

A \emph{Henneberg sequence} is a sequence of graphs
$G_2,G_3,\dots, G_{\abs{V}}$ with $G_2$ being the complete graph
on two vertices $K_2$ and each graph $G_i$ ($i\geq 3$) can be
obtained from $G_{i-1}$ by either a vertex addition operation or
an edge splitting operation. Since these operations preserve
minimal rigidity and since $K_2$ is minimally rigid, every graph
in such a sequence is minimally rigid.\\%

A simple degree counting argument shows that every minimally rigid
graph $G_{\abs{V}}=(V,E)$ with more than 2 vertices contains at
least one vertex with degree 2 or 3. One can thus always perform
either a reverse vertex addition or a reverse edge splitting
operation and obtain a smaller minimally rigid graph
$G_{\abs{V}-1}$. Doing this recursively, one eventually obtains a
minimally rigid graph on two vertices, which can only be $K_2$. It
is straightforward to see that the sequence $K_2 = G_2, G_3,\dots
G_{\abs{V}}$ is then a Henneberg sequence. We have thus proved the
following result \cite{TayWhiteley:85}:

\begin{thm}\label{thm:undirerected_Henneberg}
Every minimally rigid graph on more than one vertex can be
obtained as the result of a Henneberg sequence.
\end{thm}

The result of Theorem \ref{thm:undirerected_Henneberg} provides a
way to exhaustively enumerate all minimally rigid graphs. One can
thus use it to obtain an upper bound on the number of minimally
rigid graph having a certain number of vertices. However, this
only provides an upper bound for the Henneberg sequence allowing
one to build a certain minimally rigid graph is usually not
unique. The graph in Figure \ref{fig:rigidity}(b) can for example
be obtained from $K_2$ by either two vertex additions or one
vertex addition followed by one edge splitting.

\subsection{Minimally persistent graphs}\label{sec:min_per}

Consider a group of autonomous agents represented by vertices of a
graph. To each of these agents, one assigns a (possibly empty) set
of unilateral distance constraints represented by directed edges:
the notation $(i,j)$ for a directed edge connotes that the agent
$i$ has to maintain its distance to $j$ constant during any
continuous move. The persistence of the directed graph means that
provided that each agent is trying to satisfy its constraints, the
distance between any pair of connected or non-connected agents is
maintained constant during any continuous move, and as a
consequence the shape of the formation is preserved. A formal
definition of persistence is
given in \cite{HendrickxAndersonDelvenneBlondel:2005}.\\%

In a two-dimensional space, an agent having only one distance
constraint to satisfy can move on a circle centered on its
neighbor, and has thus one degree of freedom. Similarly, an agent
having no distance constraint to satisfy can move freely in the
plane and has thus two degrees of freedom. We call the
\emph{number of degrees of freedom} of a vertex $i$ the (generic)
dimension of the set in which the corresponding agent can chose
its position (all the other agents being fixed). It represents
thus in some sense the decision power of this agent. The number of
degrees of freedom of a vertex $i$ is given by
$\max\prt{0,2-d^+(i)}$ (where $d^+(i)$ and $d^-(i)$ represent
respectively the in- and out-degree of the vertex $i$).\\%

A graph is \emph{minimally persistent} if it is persistent and if
no edge can be removed without losing persistence. The following
result provides a swift criterion to decide minimal persistence :
\begin{prop}{\cite{HendrickxAndersonDelvenneBlondel:2005}}\label{prop:minper}
A graph is minimally persistent if and only if it is minimally
rigid and no vertex has an out-degree larger than 2.
\end{prop}

As a consequence of Proposition \ref{prop:minper}, the number of
degrees of freedom of a vertex $i$ in a minimally persistent graph
is $2-d^+(i)$. By Proposition \ref{prop:Laman_min}, it follows
after summation on all the vertices that the total number of
degrees of freedom present in such a graph is always 3. This
result is consistent with the intuition, there are indeed three
degree of freedom to chose the position and orientation of a rigid
body in a 2-dimensional space.\\%

\begin{figure}
\centering
\begin{tabular}{c}
\includegraphics[scale=.25]{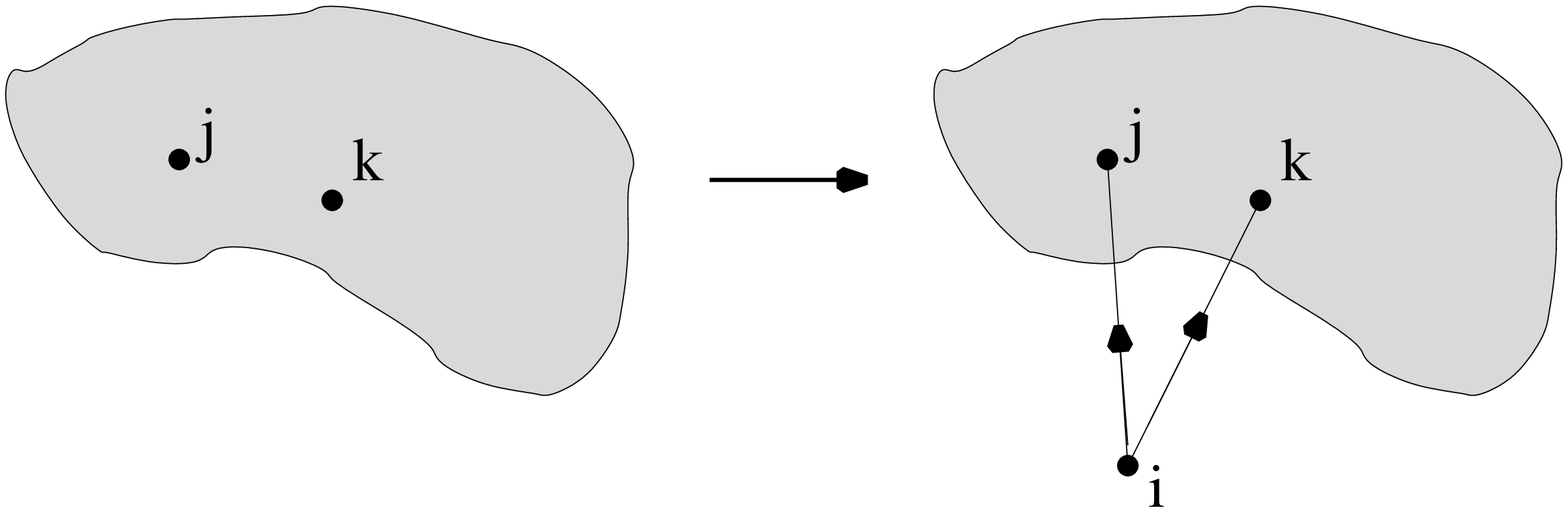}\\
(a)\\
\includegraphics[scale=.25]{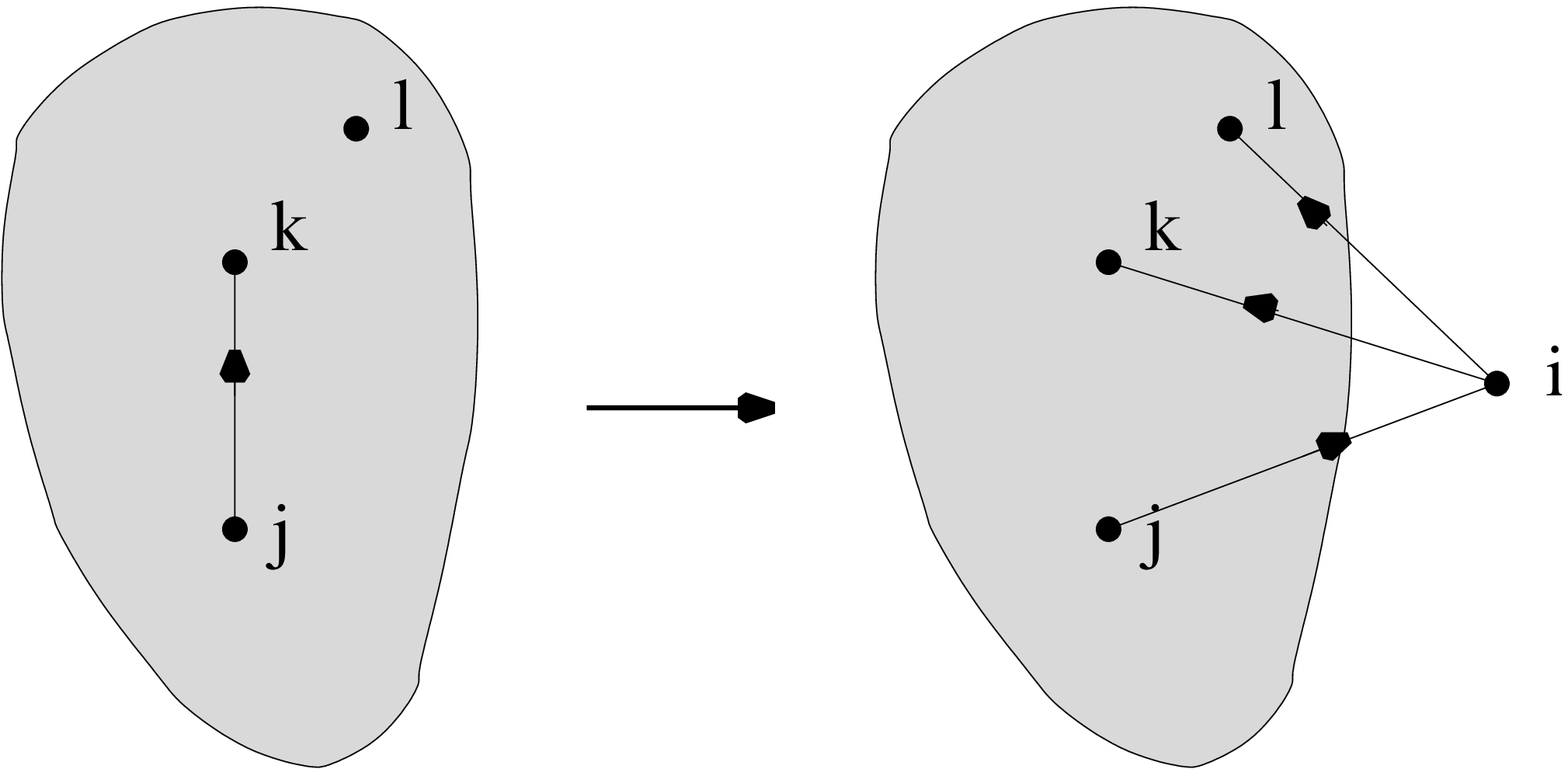}
\\
(b)\\
\end{tabular}
\caption{Representation of the directed vertex addition (a) and
edge splitting (b).}\label{fig:repres_dir_op}
\end{figure}

Let $j,k$ be two distinct vertices of a minimally persistent graph
$G=(V,E)$. A \emph{directed vertex addition}
\cite{ErenWhiteleyAndersonMorseBelhumeur:2005,HendrickxAndersonBlondel:2005}
consists in adding a vertex $i$ and two directed edges $(i,j)$ and
$(i,k)$ as shown in Figure \ref{fig:repres_dir_op}(a). Since it is
a vertex addition operation, it preserves minimal rigidity.
Besides, the added vertex has an out-degree 2 and the out-degree
of the already existing vertices are unchanged. By Proposition
\ref{prop:minper}, the directed vertex addition thus preserves the
minimal persistence. Moreover, if a vertex has an out-degree 2 and
an in-degree 0 in a minimally persistent graph, one can always
perform a reverse (directed) vertex addition by removing it, and
obtain a smaller minimally persistent graph.\\%

Let $(j,k)$ be a directed edge in a minimally persistent graph and
$l$ a distinct vertex. A \emph{directed edge splitting}
\cite{ErenWhiteleyAndersonMorseBelhumeur:2005,HendrickxAndersonBlondel:2005}
consists in adding a vertex $i$, an edge $(i,l)$, and replacing
the edge $(j,k)$ by $(j,i)$ and $(i,k)$, as shown in Figure
\ref{fig:repres_dir_op}(b). Again, this operation preserves
minimal rigidity since it is an edge splitting operation from an
undirected point of view, and since the added vertex has an
out-degree 2 and the out-degree of the already existing vertices
are unchanged, it also preserves minimal persistence. But, unlike
in the case of directed vertex addition, \emph{the reverse
operation cannot always be performed}. Suppose indeed that we have
a vertex $i$ with out-degree 2 and in-degree 1, and call its
neighbors $j,k$ and $l$. The reverse operation consists in
removing $i$ and its incident edges, and adding either $(j,k)$ or
$(j,l)$ (note that $k$ and $l$ are interchangeable). Adding any
other edge such as $(k,l)$ or $(l,k)$ would indeed prevent the
operation from being out-degree preserving, and one could then not
guarantee the minimal persistence of the graph obtained (by
Theorem \ref{prop:minper}). But, it can happen that both pairs
$(j,l)$ and $(j,k)$ are already connected by explicit or implicit
edges. In such a case, minimal rigidity is only preserved by
addition of an edge between $k$ and $l$, which as explained above
may not preserve persistence.

\begin{figure}
\centering
\includegraphics[scale=.3]{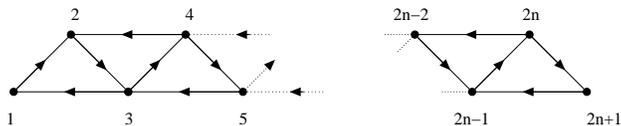}
\caption{Class of graphs on which no reverse vertex addition or
edge splitting can be performed.}\label{fig:infinity_no_reverse}
\end{figure}

We now show that the vertex addition and edge splitting operations
do not allow one to grow all minimally persistent graphs from an
initial seed. Consider the graph in Figure
\ref{fig:infinity_no_reverse} (for $n\geq 1$). One can verify by
Theorem \ref{thm:undirerected_Henneberg} that it is minimally
rigid. Moreover, no vertex has an out-degree larger than 2; by
Proposition \ref{prop:minper} it is thus minimally persistent.
Observe that no vertex has an in-degree 0; it is thus impossible
to perform a reverse vertex addition operation. Moreover, only the
vertex $2n$ satisfies the required conditions about the in- and
out-degree in order to offer the possibility of removal by a
reverse edge splitting operation, and one can verify that this
operation cannot be performed due to the presence in the graph of
the edges $(2n+1,2n-1)$ and $(2n-2,2n-1)$. Since this is true for
any $n$, we have an infinite class of graphs on which none of the
two above defined reverse operations can be performed (Note that
there provably exists other such infinite classes in which the
graphs have one vertex with two degrees of freedom and one vertex
with one degree of freedom instead of three vertices with one
degree of freedom). As a consequence, it is not possible to build
every minimally persistent graph by performing a sequence of
directed vertex addition or edge splitting operations on some seed
graph taken in a finite set of graphs. However, we have the
following less powerful result (as argued in
\cite{HendrickxAndersonDelvenneBlondel:2005}).

\begin{prop}\label{prop:1minper_to_each_minrig}
It is possible to assign directions to the edges of any minimally
rigid graph such that the obtained directed graph is minimally
persistent and can be obtained by performing a sequence of vertex
additions and edge splittings on an initial graph of two vertices
connected by one directed edge (called a \quotes{leader-follower
seed}).
\end{prop}
\begin{proof}
Let $G$ be a minimally rigid (undirected) graph. By Theorem
\ref{thm:undirerected_Henneberg}, it can be obtained by performing
a sequence of undirected vertex additions and edge splittings on
$K_2$. By performing the same sequence of the directed version of
these operations on an initial leader-follower seed, one obtains a
directed graph having $G$ as underlying undirected graph.
Moreover, since this initial seed is trivially minimally
persistent (by Proposition \ref{prop:minper}), and since the
directed versions of both vertex addition and edge splitting
preserve minimal persistence, the obtained graph is minimally
persistent.
\end{proof}

In the following sections, we examine different possibilities of
additional operations that allow the construction of all minimally
persistent graphs. In order to avoid confusion, we shall sometimes
refer to the directed version of vertex addition and edge
splitting as \emph{standard vertex addition} and \emph{standard
edge splitting}. We denote by $\S$ the set consisting of these two
operations and $\S^{-1}$ the one consisting of their inverses (the
same convention is used in the sequel for all the operations set).
Note that it is always possible to perform an operation of $\S$ on
a minimally persistent graph, but we have seen that this is not
true for operations of $\S^{-1}$.

\section{A purely directed operation}\label{sec:third_op}

We introduce here a third persistence-preserving operation:
\emph{the edge reversal}. Unlike those of $\S$, does not affect
the underlying undirected graph. We then define to
macro-operations which help us to prove that the edge reversal is
sufficient to obtain any minimally persistent graph from any other
one having the same underlying undirected graph, and show how this
implies that this operation combined with those of $\S$ is
sufficient to obtain any minimally persistent graph from a unique
initial seed.

\subsection{Edge reversal}\label{sec:edge_reversal}

Let $(i,j)$ be an edge such that $j$ as at least one degree of
freedom, i.e., $d^+(j)=0$ or $d^+(j)=1$. The \emph{edge reversal}
operation consists in replacing the edge $(i,j)$ by $(j,i)$. As a
consequence, one degree of freedom is transferred from $j$ to $i$.
This operation is its auto-inverse and preserves minimal
persistence since it does not affect the underlying undirected
graph and the only increased out-degree $d^+(j)$ remains no
greater than 2. From an autonomous agent point of view $j$
transfers its decision power (or a part of it) to $i$.

\subsection{Path reversal}

Given a directed path $P$ between a vertex $i$ and a vertex $j$
such that $j$ has a positive number of degrees of freedom, a
\emph{path reversal} consists in reversing the directions of all
the edges of $P$. As a result, $j$ loses a degree of freedom, $i$
acquires one, and there is a directed path from $j$ to $i$.
Moreover, the number of degrees of freedom of all the other
vertices remain unchanged. Note that $i$ and $j$ can be the same
vertex, in which case the path either has a trivial length $0$ or
is a cycle. In both of these situations, the number of degrees of
freedom is preserved for every vertex.\\%

\begin{figure}
\centering
\begin{tabular}{c}
\includegraphics[scale=.4]{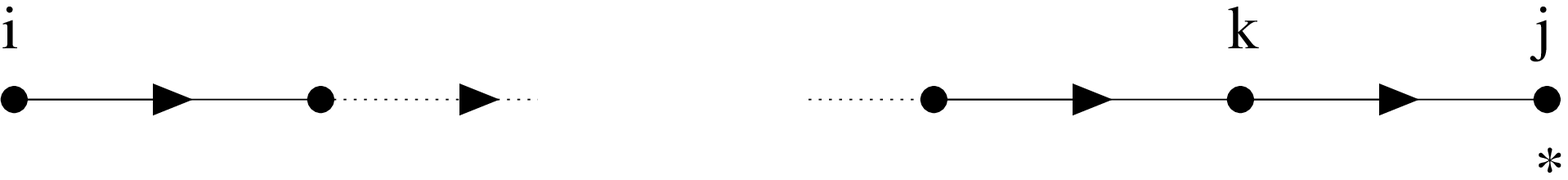}\\\\
\includegraphics[scale=.4]{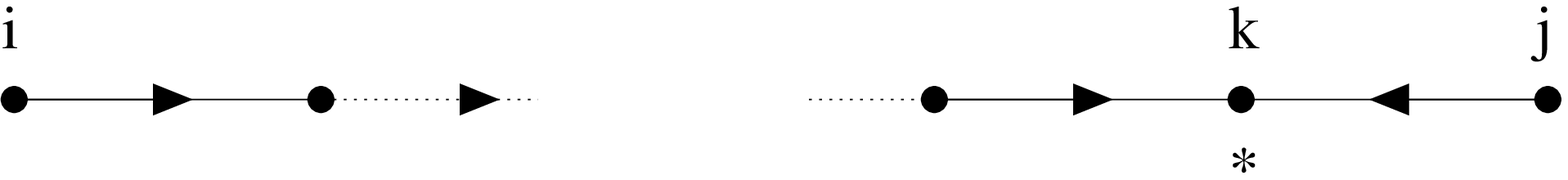}\\\\
\includegraphics[scale=.4]{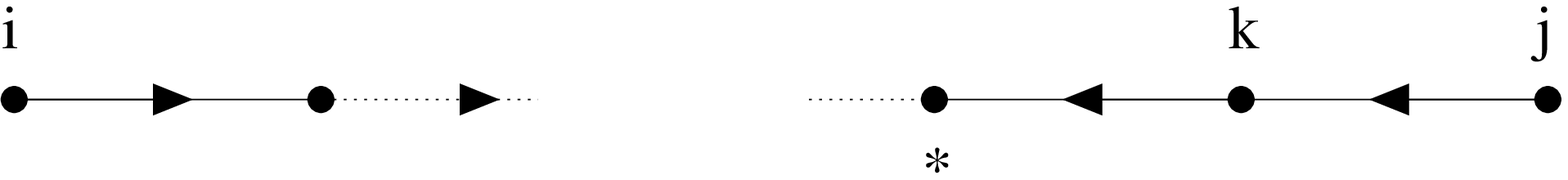}\\\\
\includegraphics[scale=.4]{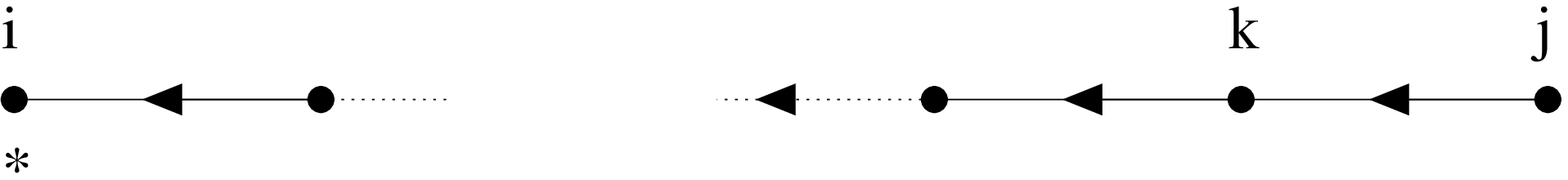}
\end{tabular}
\caption{Implementation of the path reversal by a sequence of edge
reversals. The symbol \quotes{*} represents one degree of
freedom.}\label{fig:implem_path_rev}
\end{figure}

The path reversal can easily be implemented with a sequence of
edge reversals: Since $j$ has a degree of freedom, one can reverse
the last edge of the path, say $(k,j)$, such that $j$ loses one
degree of freedom while $k$ acquires one. One can then iterate
this operation along the path until $i$, as shown in Figure
\ref{fig:implem_path_rev}. At the end, $i$ has have an additional
degree of freedom, $j$ has lost one, and all the edges of the path
have have been reversed. Note that the sequence of edge reversals
can usually not be performed in another order, for the condition
about the availability of a degree of freedom would not be
satisfied. The final result would be the same, but all the
intermediate graphs would not necessarily be minimally
persistent.\\%

We now show that this operation allows one to reposition the three
degrees of freedom of a minimally persistent graph onto any chosen
vertices (with at most two degrees of freedom on a single vertex).
For this purpose, we need the following result (which is a
particular case of a result available in
\cite{YuHendrickxFidanAndersonBlondel:2005,
HendrickxFidanYuAndersonBlondel:2005_conf}):

\begin{prop}\label{prop:path_to_dof}
Let $G$ be a minimally persistent graph, $i$ and $j$ two vertices
of $G$ with $d^+(i)\geq 1$ and $d^+(j)\leq 1$. Then, there is a
directed path from $i$ to $j$.
\end{prop}
\begin{proof}
Suppose (to obtain a contradiction) that we have a minimally
persistent (and thus minimally rigid) graph $G=(V,E)$, a vertex
$i$ with positive out-degree and a vertex $j$ with a positive
number of degrees of freedom such that there is no directed path
connecting $i$ to $j$. Let $V''\subset V$ be the set of vertices
that can be reached from $i$, and $E''$ the set of edges that
leave vertices of $V''$. Obviously, every edge of $E''$ is
incident to two vertices of $V''$, and because $d^+(i)>0$, we have
$\abs{V''}\geq 2$ and $\abs{E''}\geq 1$. Moreover, the sum on the
vertices of $V''$ of the numbers of degrees of freedom (which we
denote $F(V'')$) is smaller than $3$. There are indeed only three
degrees of freedom in a minimally persistent graph as explained in
Section \ref{sec:min_per}, and the vertex $j$ which has at least
one of them does by hypothesis not belong to $V'$. Since every
vertex has an out-degree smaller no greater than two in a
minimally persistent graph (by Proposition \ref{prop:minper}), we
have thus a subgraph $G''=(V'',E'')$ such that
\begin{equation*}
\abs{E''} = \sum_{k\in V''} d^+(k,G'')= 2 \abs{V''}-F(V'')
> 2\abs{V''}-3,
\end{equation*}
which by Proposition \ref{prop:Laman_min} is impossible for a
subgraph of a minimally rigid graph.
\end{proof}

Let us now suppose that one wants to transfer a degree of freedom
from some vertex $j$ to some vertex $i$ which has at most one
degree of freedom (transferring a degree of freedom to a vertex
that has already two degrees of freedom would indeed be impossible
as there is no edge inwardly incident). By Proposition
\ref{prop:path_to_dof}, there exists a directed path from $i$ to
$j$. The transfer can then be done by reversing this path, which
leaves the positions of all the other degrees of freedom
unchanged. By doing this at most three times, one can thus
reposition the three degrees of freedom onto any chosen vertices.
As a consequence, we have the following result.

\begin{prop}\label{prop:positionning_dof}
Let $G_A$ and $G_B$ be two minimally persistent graphs having the
same underlying undirected graph. By applying a sequence of at
most three path reversals on $G_A$, it is possible to obtain a
minimally persistent graph $G_A'$ in which every vertex has the
same number of degrees of freedom as in $G_B$
\end{prop}

\subsection{Cycle reversal}

A \emph{cycle reversal} consists in reversing all the edges of a
directed cycle. This operation does not affect the number of
degrees of freedom of any vertex nor the underlying
undirected graph, and preserves therefore minimal persistence.\\%

\begin{figure}
\centering
\begin{tabular}{cccc}
\includegraphics[scale=.3]{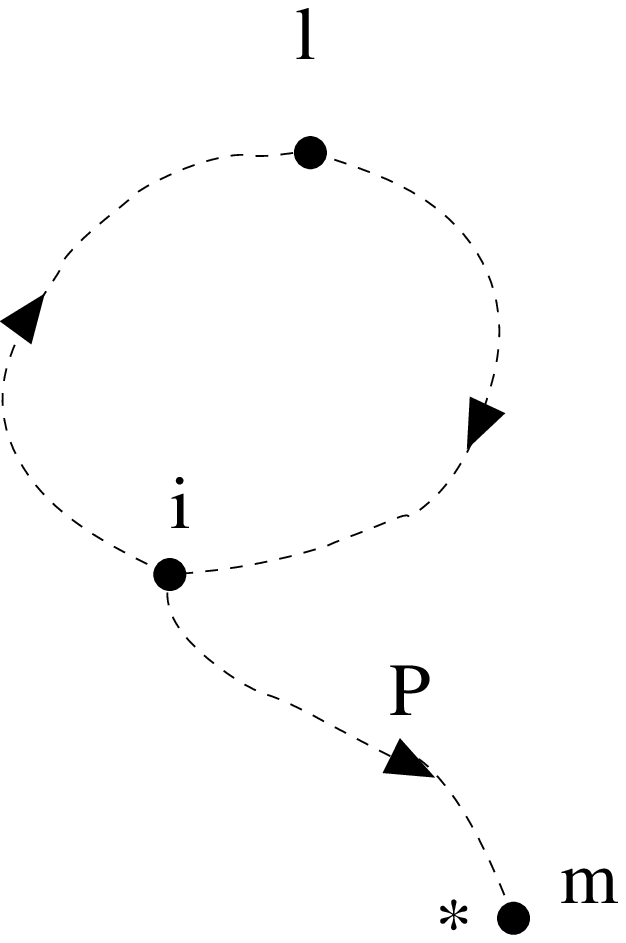}&
\includegraphics[scale=.3]{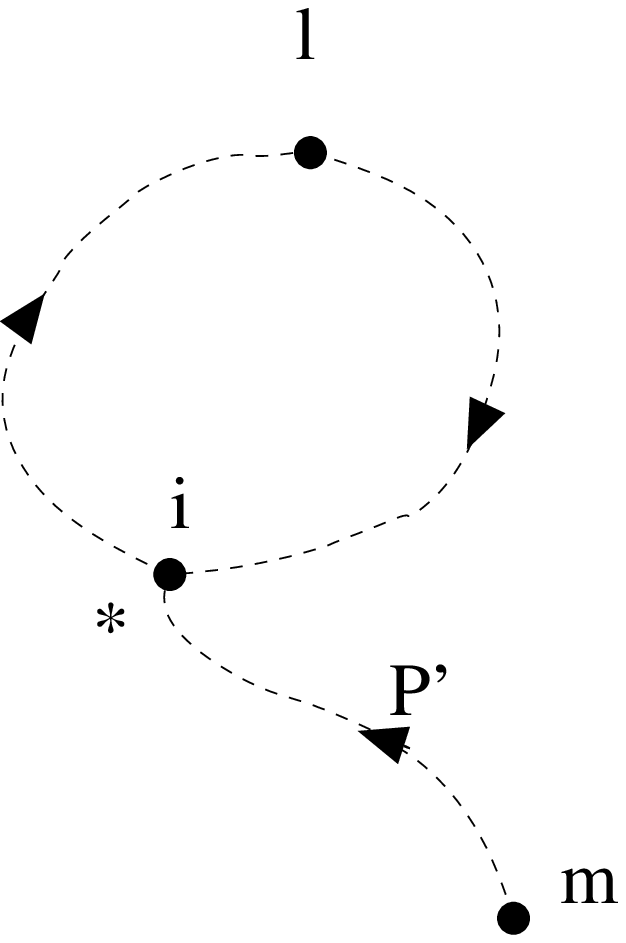}&
\includegraphics[scale=.3]{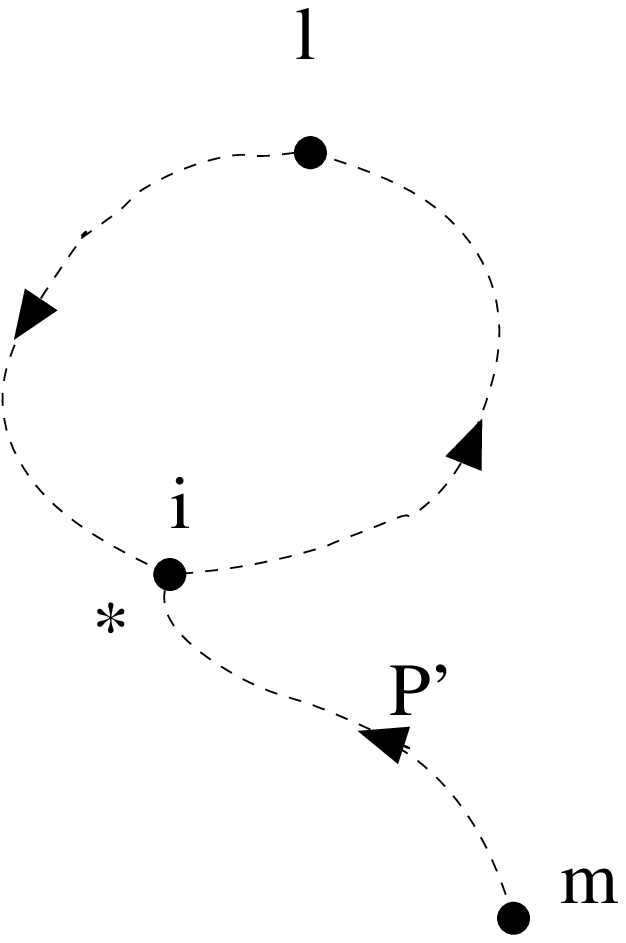}&
\includegraphics[scale=.3]{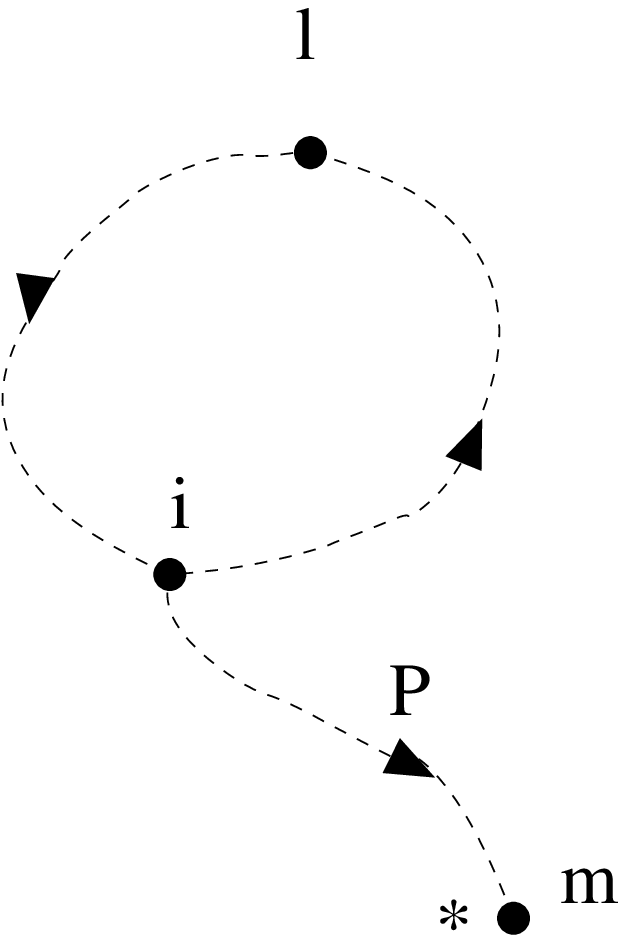}
\end{tabular}
\caption{Implementation of cycle reversal. The \quotes{*}
represents one degree of freedom.}\label{fig:implem_cycle_rev}
\end{figure}

A cycle reversal on a minimally persistent graph can be
implemented by a sequence of edge reversals. Let us indeed first
suppose that there is a vertex $i$ in the cycle that has at least
one degree of freedom. In that case, the cycle reversal is just a
particular case of the path reversal, with $i=j$. We now assume
that no vertex in the cycle has a degree of freedom. Let $l$ be a
vertex in the cycle, and $m$ a vertex that does not belong to the
cycle but has a degree of freedom. By Proposition
\ref{prop:path_to_dof}, it follows that there exists a directed
path from $l$ to $m$. Let $i$ be the last vertex in this path
belonging to the cycle. There is trivially a path $P$ from $i$ to
$m$ such that every other vertex of this path does not belong to
the cycle. The implementation of a cycle reversal by three path
reversals is then represented in Figure
\ref{fig:implem_cycle_rev}. One begins by reversing the path $P$
into $P'$ such that $i$ acquires a degree of freedom. As explained
above, the cycle can then be reversed since it is a particular
case of path reversal, and finally, one reverses the path $P'$
back to $P$ such that the degree of freedom acquired by $i$ is
re-transmitted to $m$. Note that an alternative equivalent
approach is to reverse the path from $l$ to $m$ containing $i$ and
one part of the cycle, and then to reverse the newly created path
from $m$ to $l$ containing $i$ and the other part of the cycle.

\begin{rem} Both cycle reversal and path reversal are
their auto-inverse, as is the case for edge reversal. Moreover,
the fact that they can be implemented using only edge reversals is
another way to show that they preserve minimal persistence.
\end{rem}

We now prove that from any minimally persistent graph, one can
obtain any other minimally persistent graph having the same
underlying undirected graph and allocation of degrees of freedom
by a sequence of cycle reversals. For this purpose, we need the
following result.

\begin{lem}\label{lem:cycle_opposite_edges}
Let $G_A=(V,E_A)$ and $G_B=(V,E_B)$ be two graphs having the same
underlying undirected graph and such that every vertex has the
same out-degree in both graphs.  If an edge of $G_A$ has the
opposite direction to that in $G_B$, it belongs to (at least) one
cycle of such edges in $G_A$.
\end{lem}
\begin{proof}
Suppose that $(i_0,i_1)\in E_A$ and $(i_1,i_0)\in E_B$ (i.e., this
edge has opposite directions in $G_A$ and $G_B$); then there
exists at least one vertex $i_2\not=i_0$ such that $(i_1,i_2)\in
E_A$ and $(i_2,i_1) \in E_B$. For if the contrary holds, we would
have $d^+(i_1,G_A) = d^+(i_1,G_B) -1$, which contradicts our
hypothesis. Repeating this argument recursively, we obtain an
(infinite) sequence of vertices $i_0,i_1,i_2,\dots$ such that for
each $j\geq 0$, $(i_j,i_{j+1}) \in E_A$ and $(i_{j+1},i_{j}) \in
E_B$. Since there are only a finite number of vertices in $V$, at
least one of them will appear twice in this sequence. By taking
the subsequence of vertices (and induced edges) appearing in the
infinite sequence between any two of its occurrences we obtain
then a cycle having the required properties.
\end{proof}

\begin{prop}\label{prop:reversing_cycles}
Let $G_A=(V,E_A)$ and $G_B=(V,E_B)$ be two minimally persistent
graphs having the same underlying undirected graph and such that
every vertex has the same number of degrees of freedom in both of
them. Then it is possible to obtain $G_B$ from $G_A$ by a sequence
of at most $\abs{E_A}/3=\abs{E_B}/3$ cycle reversals.
\end{prop}
\begin{proof}
Suppose that $G_A\not=G_B$, and let $E_o$ denote the set of edges
of $G_A$ that do not have the same direction as in $G_B$. Since
both graphs have the same underlying undirected graph and since
all the vertices have the same out-degrees in both of them it
follows from Lemma \ref{lem:cycle_opposite_edges} that there
exists a cycle of edges of $E_o$. $\abs{E_o}$ is thus strictly
decreased by reversing this cycle. Doing this recursively leads
then to $\abs{E_o} =0$, i.e, to a situation where $G_A = G_B$.
Moreover, since every cycle has at least three edges. The number
of cycle reversals is at most $\abs{E}/3$
\end{proof}

\subsection{Obtaining all minimally persistent graphs using
three primitive operations}\label{sec:obtention_with_T}

Using the results of the two previous subsections, we can now show
the following Proposition.
\begin{prop}\label{prop:class=undir_struc}
By applying a sequence of edge reversals to a given minimally
persistent graph, it is possible to obtain any other minimally
persistent graph having the same underlying undirected graph.
Moreover, all the intermediate graphs are then minimally
persistent.
\end{prop}
\begin{proof}
Since both path reversal and cycle reversal can be implemented by
a sequence of edge reversals (which preserves minimal
persistence), this result is a direct consequence of Propositions
\ref{prop:positionning_dof} and \ref{prop:reversing_cycles}.
\end{proof}
From an autonomous agent formation perspective, suppose that a
reorganization of the distance constraints distribution needs to
be performed, and that this reorganization preserves the structure
of constraints from an undirected point of view, i.e., the
reorganization is solely one involving changes of some directions.
Proposition \ref{prop:class=undir_struc} implies that this can be
done by a sequence of local degree of freedom transfers, in such a
way that during all the intermediate stages, the formation shape
is guaranteed to be maintained as a result of persistence being
preserved.\\%

Let $\T$ be the set of operations containing vertex addition, edge
splitting, and edge reversal. A \emph{leader-follower seed} is a
minimally persistent graph on two vertices. It contains only one
edge, leaving a vertex called \quotes{the follower}, and arriving
at the other one, called \quotes{the leader}. The next theorem
states that one can obtain any minimally persistent graph from an
initial leader-follower seed using only operations of $\T$.

\begin{thm}\label{thm:3op}
Every minimally persistent graph can be obtained by applying a
sequence of operations of $\T$ to an initial leader-follower seed.
Moreover, all the intermediate graphs are minimally persistent.
\end{thm}
\begin{proof}
Consider a minimally persistent graph $G$. This graph is also
minimally rigid. By Proposition \ref{prop:1minper_to_each_minrig},
there exists thus a (possibly different) minimally persistent
graph having the same underlying undirected graph that can be
obtained by performing a sequence of operations of $\S\subset \T$
on an initial leader follower seed. By Proposition
\ref{prop:class=undir_struc}, $G$ can then be obtained by applying
a sequence of edge reversals on this last graph. Moreover, since
all the operations of $\T$ preserve minimal persistence, all the
intermediate graphs are minimally persistent.
\end{proof}

\begin{figure}
\centering
\begin{tabular}{c}
\includegraphics[scale =.3]{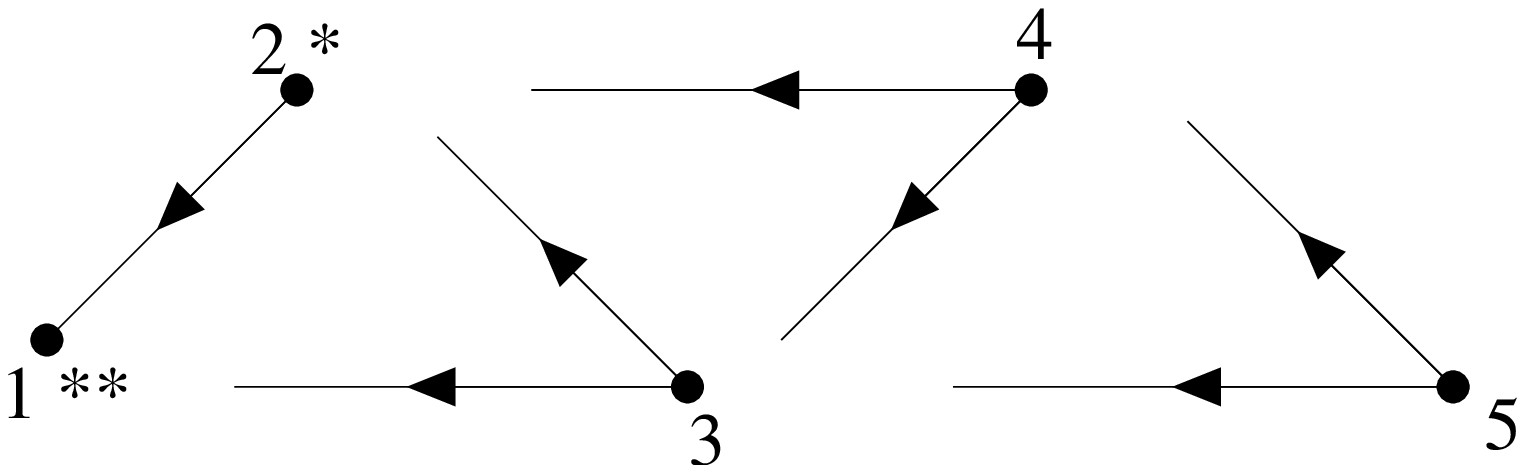}\\\\
(a)\\\\
\begin{tabular}{l}
\includegraphics[scale =.3]{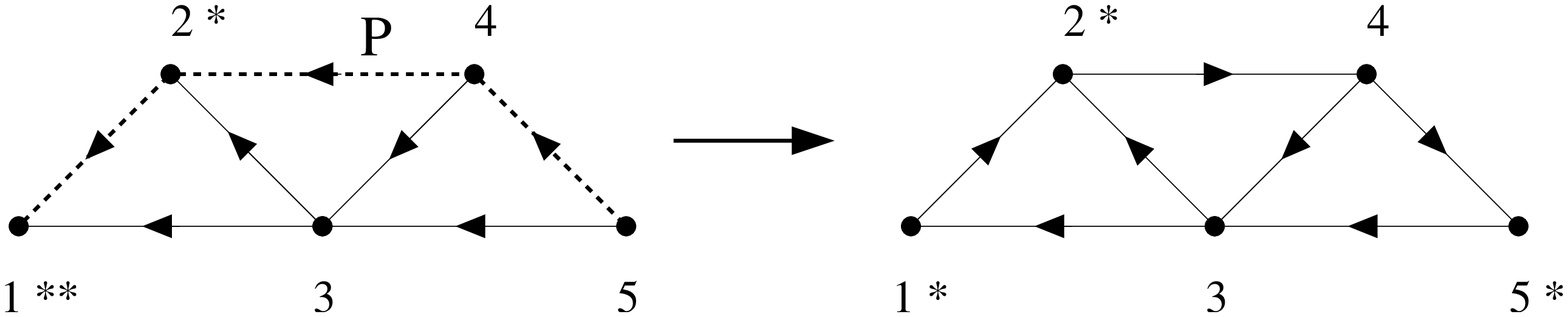}
\end{tabular}\\
(b)\\

\begin{tabular}{l}
\includegraphics[scale =.3]{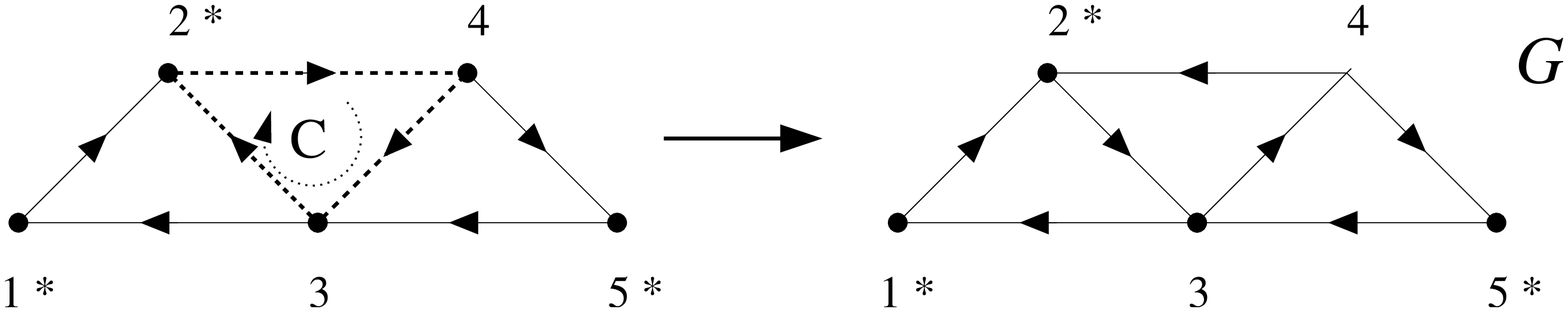}
\end{tabular}\\
(c)
\end{tabular}

\caption{Example of obtaining of a minimally persistent graph by
applying a sequence of operations of $\T$ on a leader-follower
seed. The graph $G$ is obtained from the leader-follower seed by
(a) three vertex additions, (b) the reversal of the path $P$ and
(c) of the cycle $C$. }\label{fig:example_use_T}
\end{figure}

As an illustration of Theorem \ref{thm:3op}, consider the graph
$G$ represented in the right hand side of Figure
\ref{fig:example_use_T}(c), which is an instantiation of the graph
of Figure \ref{fig:infinity_no_reverse} with $n=2$. As explained
in Section \ref{sec:min_per}, it cannot be obtained by applying a
vertex addition or an edge splitting on a smaller minimally
persistent graph. However, by Theorem \ref{thm:3op}, it can be
obtained by applying a sequence of operations of $\T$ on an
initial leader-follower seed. Let us take $1$ and $2$ as
respectively leader and follower of this initial seed. One can
begin by adding $3$, $4$ and $5$ using three vertex additions as
shown in Figure \ref{fig:example_use_T}(a). The graph obtained has
the same underlying undirected graph as $G$, but the degrees of
freedom are not allocated to the same vertices. By reversing the
path $P$ ($\{5,4,2,1\}$) (using a sequence of edge reversals), one
can then transfer one degree of freedom from $1$ to $5$ as shown
in Figure \ref{fig:example_use_T}(b) such that in the obtained
graph, all the vertices have the same number of degrees of freedom
as in $G$. As stated in Proposition \ref{prop:reversing_cycles},
any edge of this graph that does not have the same direction as in
$G$ belongs to a cycle of such edges. The only such cycle here is
$C$. By reversing it (using a sequence of edge reversals), one
finally obtains the graph $G$, as shown in Figure
\ref{fig:example_use_T}(c). Note that consistently with Theorem
\ref{thm:3op}, all the intermediate graphs are minimally
persistent.

Theorem \ref{thm:3op} also proves that it is always possible to
obtain a leader-follower pair from any minimally persistent graph
by applying an appropriate sequence of operations of $\T^{-1}$.
This can be also stated as follows:
\begin{thm}
Let $G$ be a minimally persistent graph. By applying a (possibly
empty) sequence of edge reversals on $G$, it is always possible to
obtain a minimally persistent graph on which at least one
operation of $\S^{-1}$ (i.e, reverse edge splitting or reverse
vertex addition) can be performed.
\end{thm}

Starting from a minimally persistent graph, one can thus first use
operations of $\T^{-1}$ to obtain a leader-follower pair, and then
use operations of $\T$ to obtain any other minimally persistent
graph. This method is generally not optimal in terms of the number
of operations. However, the argument proves the following
corollary.

\begin{cor}\label{cor:tranfo_T_T^-1}
Every minimally persistent graph can be transformed into any other
minimally persistent graph using only operations of $\T\cup
\T^{-1}$.
\end{cor}

This result allows us to define a distance on the minimally
persistent graphs (on more than one vertex) by saying that the
distance between two of them is the minimal number of operations
of $\T\cup \T^{-1}$ needed to transform one into the other.
Propositions \ref{prop:1minper_to_each_minrig},
\ref{prop:positionning_dof} and \ref{prop:reversing_cycles} imply
that the distance between two graphs is quadratically bounded by
their size, the quadratic character coming from the cycle
reversing operations (the others requires only a linear number of
operations). However, a better bound is likely to exist.\\%

\begin{figure}
\centering
\includegraphics[scale = .25]{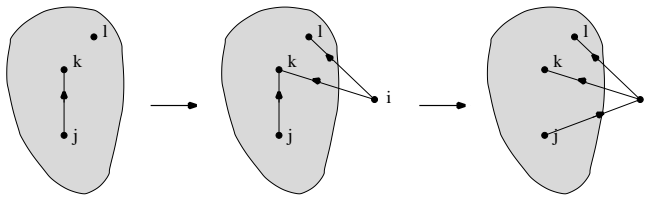}
\caption{Implementation of the edge splitting by a vertex addition
and an edge reorientation. The vertex $i$ is first added with two
out-going edges by vertex addition, and the edge $(j,k)$ is then
reoriented and becomes $(j,i)$. }\label{fig:implem_es}
\end{figure}

\begin{rem}
Observe that the three operations of $\T$ are relatively basic
ones and are performed locally. They could thus easily be
implemented in a local way on an autonomous agent formation. It
might be however possible to improve this basic character using
for example an operations such as an \emph{edge reorientation},
i.e., an operation consisting in changing the arrival vertex of an
edge. As shown in Figure \ref{fig:implem_es}, a vertex addition
operation and an edge reorientation operation can indeed implement
an edge splitting operation which could thus be discarded.
However, this would require an efficient and simple criterion to
determine when such an edge reorientation operation could be
performed.
\end{rem}

\section{An alternative set of four primitive operations} \label{sec:four_non_local}

As explained in Section \ref{sec:third_op}, every minimally
persistent graph can be obtained by applying a sequence of
operations belonging to $\T$ on an initial leader-follower seed,
in such a way that all the intermediate graphs are minimally
persistent. However, unlike in the case of an undirected Henneberg
sequence (see Section \ref{sec:Henneberg}), the number of vertices
in the final graph does not determine uniquely the required number
of intermediate graphs, but only an upper bound on it (see Section
\ref{sec:third_op}). In this section, we focus on sets of
operations equivalent to those of $\S$ from an undirected point of
view and that allow one to build all minimally persistent graphs
(the number of intermediate graphs being thus uniquely determined
by the the number of vertices of the final graph since each
operation adds one vertex). It is proved that those sets always
contain at least one operation involving the reversal of edges
that are not affected by the corresponding operation for
undirected graphs. We then provide such a set $\A$ of four types
of operations and show how it allows one to build any minimally
persistent graph $G=(V,E)$ by applying $\abs{V}-2$ operations to
an initial leader-follower seed. Finally we study the relations
between the two sets $\A$ and $\T$.

\subsection{Necessary involvement of external edges.}

In the sequel, we adopt the terms \emph{generalized vertex
addition} and \emph{generalized edge splitting} for any operation
which is equivalent to a vertex addition or an edge splitting from
an undirected point of view. Such an operation is said to be
\emph{confined} if it only affects edges that are involved in the
corresponding undirected operation. For example, all the operation
of $\S$ are confined, while the edge reversal operation defined in
Section \ref{sec:edge_reversal} is not.\\%

\begin{figure}
\centering
\includegraphics[scale= .4]{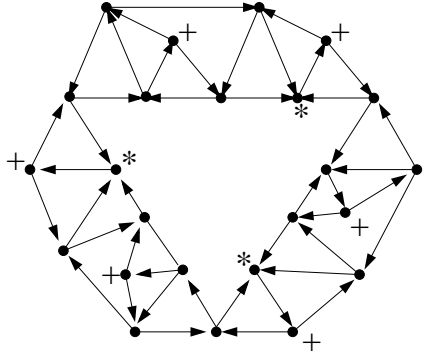}
\caption{A minimally persistent graph no vertex of which can be
removed (without losing persistence) by a reverse (generalized)
vertex addition or a confined (generalized) reverse edge
splitting. The symbol \quotes{*} represents one degree of freedom.
Vertices that are candidate to be removed by a reverse generalized
edge splitting are labelled \quotes{+}.}
\label{fig:counterex_nonlocal}
\end{figure}

Suppose that one wants to remove a vertex (without losing
persistence) from the provably minimally persistent graph
represented in Figure \ref{fig:counterex_nonlocal} using a
generalized reverse edge splitting or reverse vertex addition. The
only ones that can be removed are those with a label \quotes{+},
and due to their total degree, this could only be done by a
generalized reverse edge splitting operation. Suppose now that one
wants to use a confined version of this operation. One would then
remove one of the vertices with a label \quotes{+} and connect two
of its neighbors by a directed edge. Observe that among the three
pairs of neighbors of any vertex with a label \quotes{+}, two are
already connected, and the last pair contains two vertices with an
out-degree 2. Adding an edge between a pair of neighbors of the
removed vertex without reversing the direction of any other edge
would thus imply the presence of either a vertex with out-degree 3
(which by Theorem \ref{prop:minper} is impossible in a minimally
persistent graph) or of a cycle of length 2 (which by Proposition
\ref{prop:Laman_min} cannot appear in a minimally rigid graph).
This removal should therefore be performed by a
\emph{non-confined} reverse generalized edge splitting. The
following result is thus proved.
\begin{prop}\label{prop:nonconfined}
If a set exists of generalized vertex additions and edge
splittings allowing one to build all minimally persistent graphs
from an initial leader-follower seed, such a set must always
contain a non-confined edge splitting.
\end{prop}

The existence of confined operations that would not be equivalent
to vertex addition or edge splitting, but that would however
preserve minimal persistence and allow one to build all minimally
persistent graphs with $\abs{V}$ vertices in $\abs{V}-2$
operations (starting with a leader-follower seed) remains an open
question. Note that such operations would have to be proved to
preserve minimal rigidity.

\subsection{Description of a set $\A$ of four primitive operations}\label{sec:description_4}

We define here a new set $\A$ of four operations. The first two
are the standard vertex additions and edge splitting as described
in Section \ref{sec:min_per}, which implies that  $\S \subset \A$.
The two others are atypical versions of these.\\%

\begin{figure}
\centering
\begin{tabular}{c}
\includegraphics[scale=.25]{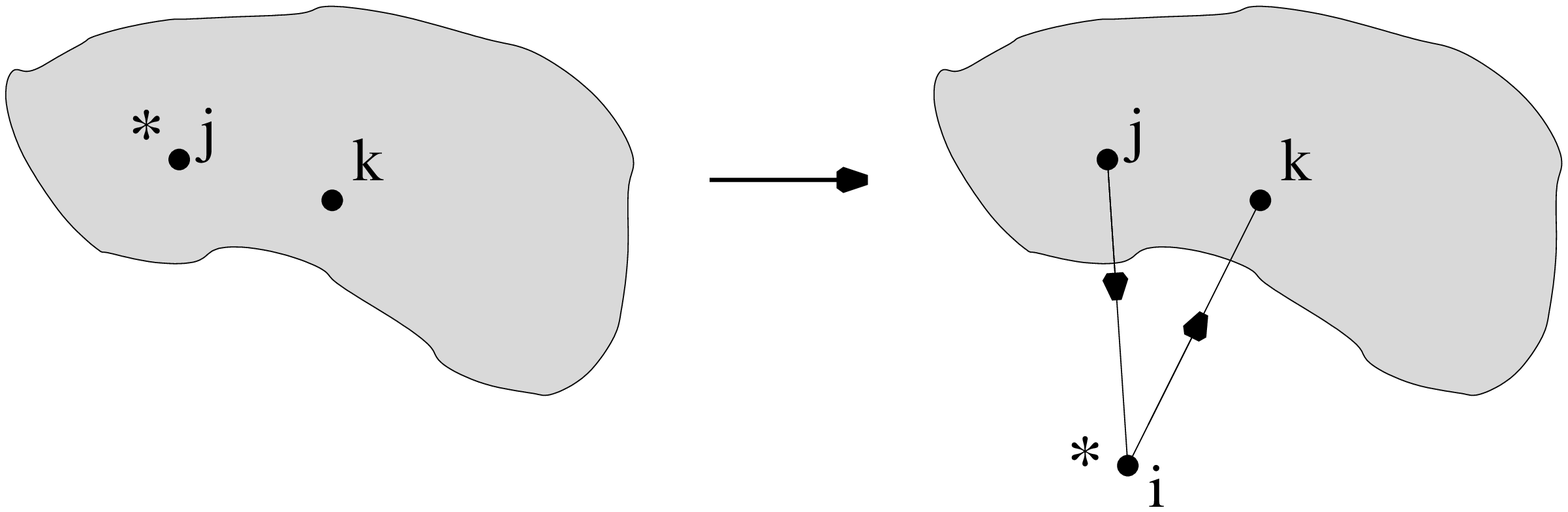}\\
(a)\\
\includegraphics[scale=.25]{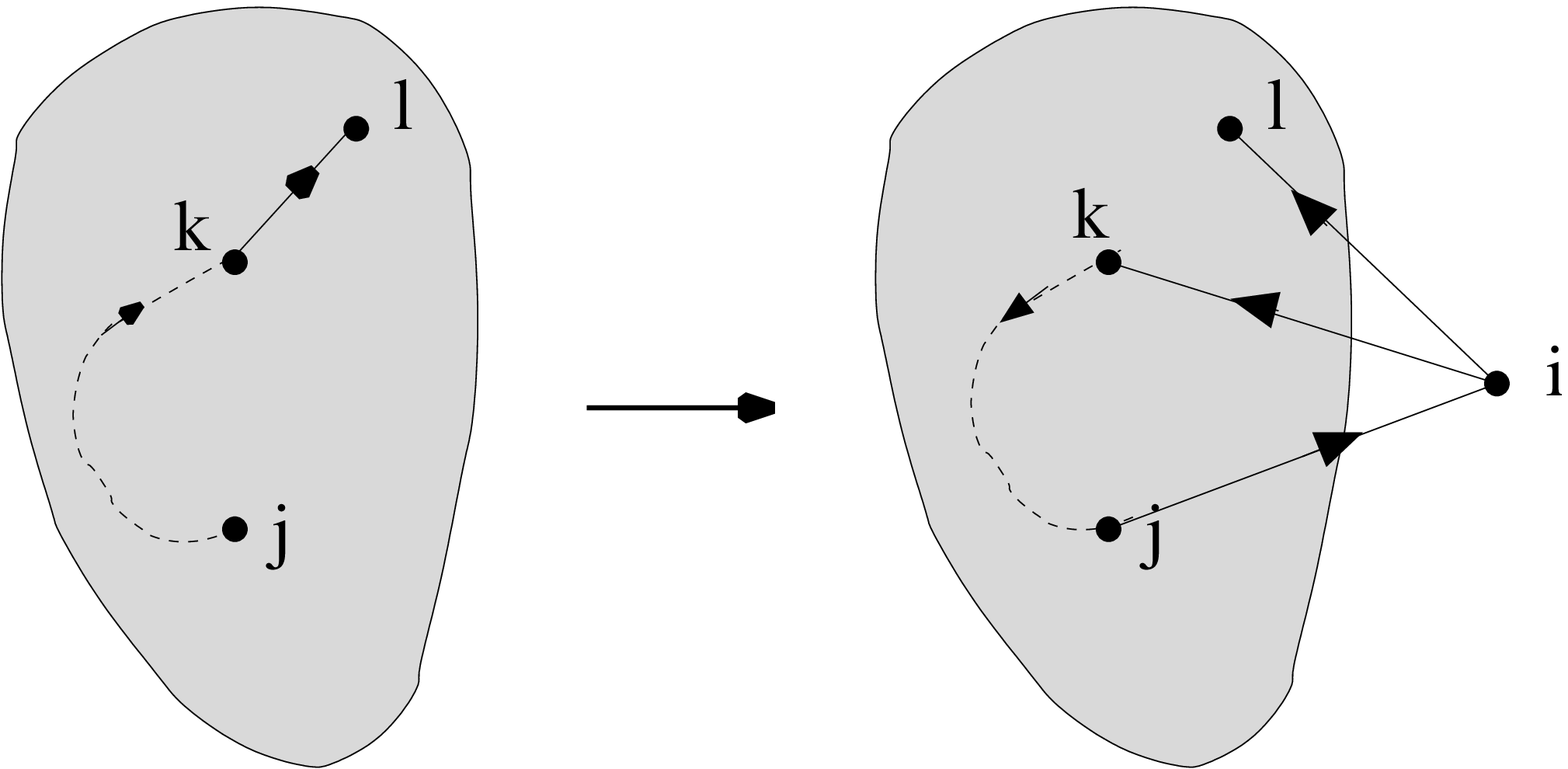}
\\
(b)\\
\end{tabular}
\caption{Representation of the atypical (a) vertex addition
operation and (b) edge splitting operation, both belonging to
$\A$. The symbol \quotes{*} represents one degree of freedom.}
\label{fig:repres_atypical operation}
\end{figure}

Let $j,k$ be two vertices of a minimally persistent graph such
that $j$ has at least one degree of freedom. The \emph{atypical
vertex addition} operation consists in adding the vertex $i$, the
edges $(j,i)$ and $(i,k)$, as shown in Figure
\ref{fig:repres_atypical operation}(a). As a result, $j$ loses a
degree of freedom, and $i$ appears with one. The \emph{reverse
atypical vertex addition} operation consists in removing a vertex
with in- and out-degree 1.

\begin{prop}
Atypical vertex addition and reverse atypical vertex addition
preserve minimal persistence.
\end{prop}
\begin{proof}
Since these operations are respectively a generalized vertex
addition and a reverse generalized vertex addition, they preserve
minimal rigidity  as explained in Section \ref{sec:min_rig_graph}.
Moreover, the reverse atypical vertex addition does not increase
the out-degree of any vertex, while the atypical vertex addition
only increases by one an out-degree that is smaller than 2. In
both situations the graph obtained after performing the operation
does not contain any vertex with an out-degree larger than 2 and
is thus minimally persistent (by Proposition \ref{prop:minper}).
\end{proof}

Let $j$, $k$ and $l$ be three vertices of a minimally persistent
graph such that there is a (simple) directed path from $j$ to $k$
and $(k,l)\in E$. The \emph{atypical edge splitting} operation
consists in removing $(k,l)$, adding a vertex $i$ and the edge
$(j,i)$, $(i,k)$ and $(i,l)$, and reversing the direction of all
the edges belonging to the path from $j$ to $k$, as represented in
Figure \ref{fig:repres_atypical operation}(b).

\begin{prop}
Atypical edge splitting preserves minimal persistence.
\end{prop}
\begin{proof}
This operation is a generalized edge splitting and thus preserves
minimal rigidity. Since it does not affect the out-degree of any
already existing vertex and adds a vertex with out-degree 2, it
also preserves minimal persistence (by Proposition
\ref{prop:minper}).
\end{proof}

Consider a vertex $i$ with out-degree 2 and in-degree 1 in a
minimally persistent graph, and call its neighbors $j$, $k$ and
$l$ as in Figure \ref{fig:repres_atypical operation}(b). Suppose
that in the graph obtained after deletion of $i$, there is a path
from $k$ (or equivalently $l$) to $j$ and the pair $(k,l)$ is not
connected by an implicit nor an explicit edge. The \emph{reverse
atypical edge splitting} consists then in removing $i$, reversing
all the edges of the path from $k$ to $j$ to obtain a path from
$j$ to $k$, and adding the edge $(k,l)$.

\begin{prop}
Reverse atypical edge splitting preserves minimal persistence.
\end{prop}
\begin{proof}
From an undirected point of view, this operation consists in
removing one vertex incident to three edges, and then connecting a
pair of unconnected vertices that does not define an implicit edge
in the intermediate graph. It thus preserves minimal rigidity.
Moreover, it does not affect the out-degree of any remaining
vertex. It follows from Proposition \ref{prop:minper} that reverse
atypical edge splitting preserve minimal persistence.
\end{proof}

The conditions in which the reverse atypical edge splitting can be
performed are not always easy to check. However, the following
result holds:

\begin{lem}\label{lem:one_among_2_reverse}
In a minimally persistent graph, a vertex with in-degree 1 and
out-degree 2 can always be removed by either a reverse standard
edge splitting or a reverse atypical edge splitting.
\end{lem}
\begin{proof}
Consider a minimally persistent graph $G=(V,E)$ and a vertex $i\in
V$ with $d^+(i)=2$, $d^-(i) =1$. We call its neighbors $j$, $k$
and $l$ such that $(j,i), (i,k), (i,l) \in E$, as in Figure
\ref{fig:one_among_2_reverse}(a)

Let us assume that $i$ cannot be removed by a reverse standard
edge splitting, i.e., that $j$ is connected to both $k$ and $l$ by
an explicit or implicit edge in $G\setminus \{i\}$. As already
mentioned in Section \ref{sec:min_rig_graph}, $k$ and $l$ are in
such a case never connected by an implicit nor an explicit edge in
$G\setminus \{i\}$, and the graph obtained by connecting them
after removing $i$ from $G$ is therefore minimally rigid. It
remains to prove the existence of a directed path from $k$ or $l$
to $j$ (note that $k$ and $l$ are interchangeable) in order that
an edge splitting operation can be applied. For this purpose, we
are going to construct a minimally persistent graph $G'$ close to
$G$ and in which $j$ has a degree of freedom. As explained below,
Proposition \ref{prop:path_to_dof} guarantees then the existence
of a directed path from either $k$ or $l$ to $j$. It will be
proved that this implies the existence of such a path in
$G\setminus \{i\}$.

Consider a vertex $p$ having at least one degree of freedom in
$G$. Since $d^+(i)=2$ in $G$, Proposition \ref{prop:path_to_dof}
guarantees the existence (still in $G$) of a (cycle-free) directed
path from $i$ to $p$. Without loss of generality, let us assume
that the second vertex of this path is $k$ (it has indeed to be a
neighbor of $i$, and $k$ and $l$ are interchangeable). There
exists thus a directed path $P$ from $k$ to $p$ to which $i$ does
not belong. We build $G'$ by reversing the path $P$ (which becomes
$P'$), removing $i$ and adding the edge $(k,l)$, as shown in
Figure \ref{fig:one_among_2_reverse}(a) and (b). As already
mentioned, any graph obtained by removing $i$ and connecting $k$
to $l$ is minimally rigid. Moreover, after the reversal of $P$ and
addition of $(k,l)$, $p$ loses one degree of freedom, i.e., its
out-degree in $G'$ is increased by one with respect to its
out-degree in $G$ (which is smaller than two). On the other hand,
$j$ acquires a degree of freedom. No vertex has thus an out-degree
larger than 2 in $G'$, which is therefore minimally persistent.

\begin{figure}
\centering
\begin{tabular}{ccc}
\includegraphics[scale=.25]{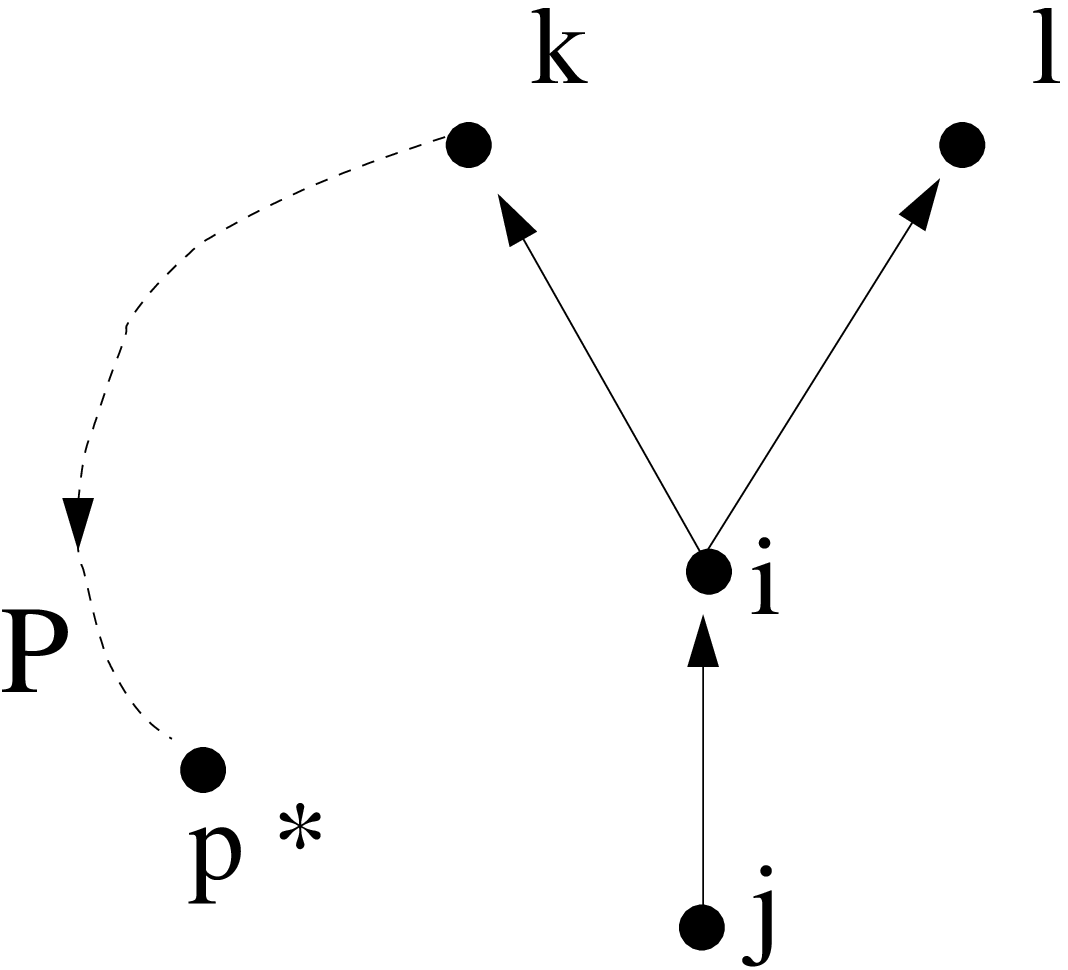}&
\includegraphics[scale=.25]{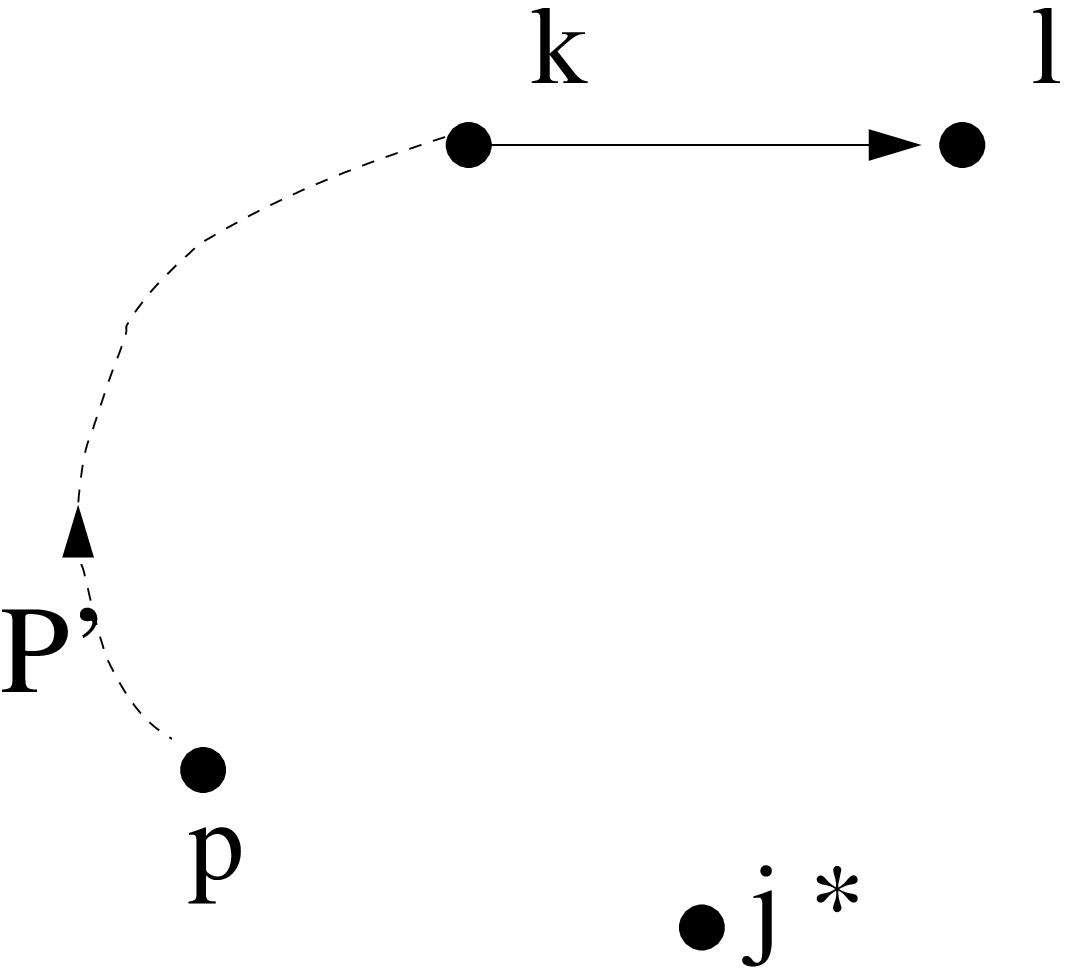}&
\includegraphics[scale=.25]{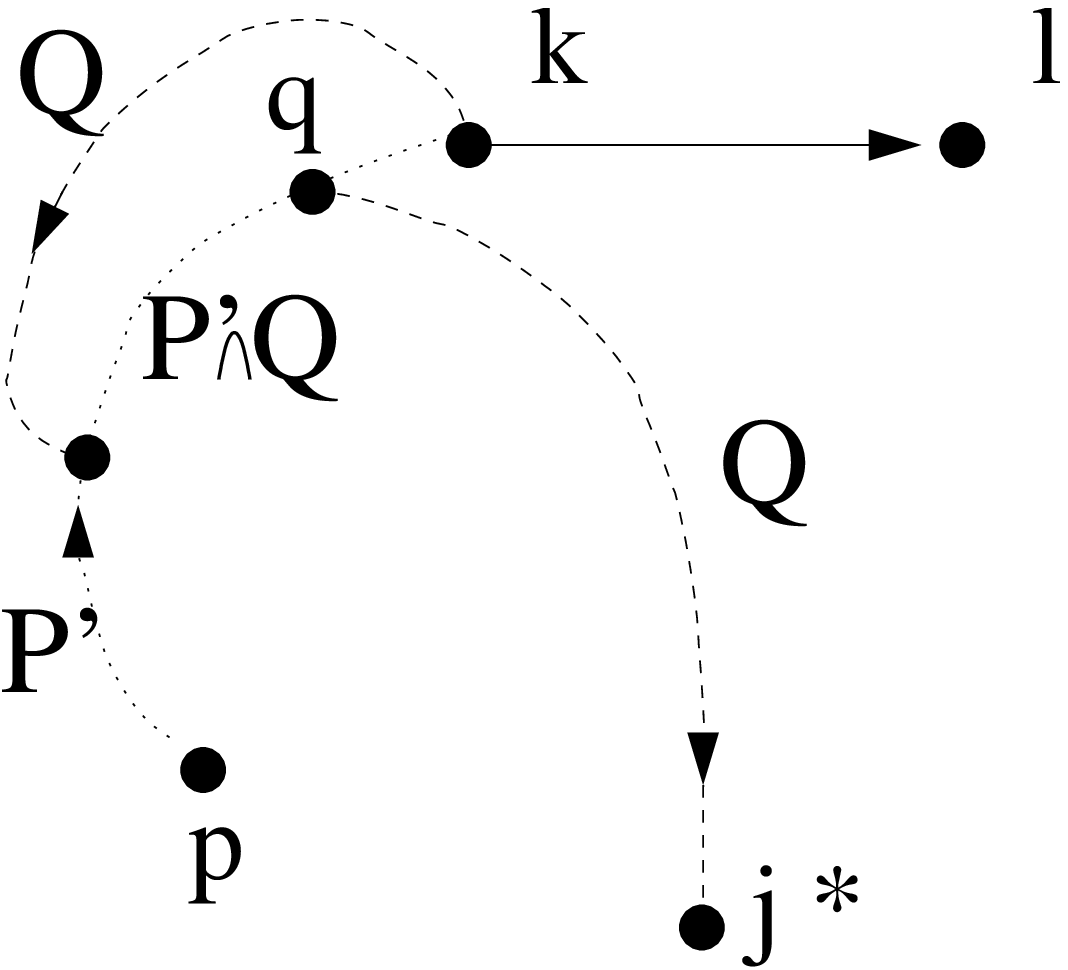}
\\&&\\
(a)&(b)&(c)
\end{tabular}

\caption{Representation of different paths and graph involved in
Lemma \ref{lem:one_among_2_reverse}. (b) shows the minimally
persistent graph $G'$ obtained from the graph $G$ represented in
(a) by removing $i$, adding $(k,l)$ and reversing $P$. (c) shows
then the path $Q$ (in $G'$) from $k$ to $j$ in a case where it has
a non empty intersection with $P'$.}
\label{fig:one_among_2_reverse}
\end{figure}

In $G'$, $k$ has by construction a positive out-degree, and $j$
has at least one degree of freedom. By Proposition
\ref{prop:path_to_dof}, there exists thus a (cycle-free) directed
path $Q$ in $G'$ from $k$ to $j$. In order to prove the existence
of such a path in $G\setminus \{i\}$, we consider three cases.
Observe that the only edges of $G'$ that do not exist in
$G\setminus \{i\}$ are $(k,l)$ and those of $P'$ (which exist but
with the opposite direction).
\begin{itemize}
\item{$Q$ and $P'$ have no common edge, $(k,l) \not \in Q$: In
that case, the path $Q$
also exists in $G\setminus \{i\}$.}%
\item{$Q$ and $P'$ have no common edge, $(k,l) \in Q$: Since the
(simple) path $Q$ does not contain any cycle, $(k,l)$ must be the
first edge of $Q$. By removing this edge, one obtains a directed
path from $l$ to $j$ having no intersection with $P'$ and that
does not contain $(k,l)$. This path exists thus also in
$G\setminus \{i\}$.}%
\item{$Q$ and $P'$ have some common edge(s): Let $q$ be the last
vertex of $Q$ that also belongs to $P'$. The edges and vertices of
$Q$ which are after $q$ constitute a directed path from $q$ to
$j$. By definition, it does not intersect $P'$, and it does not
contain $(k,l)$ for this would mean that $Q$ contains a cycle. It
exists therefore also in $G$. Moreover, since $q$ belongs to the
path $P'$ in $G'$, it belongs to the path $P$ in $G\setminus
\{i\}$, and there is thus a directed path from $k$ to $q$ in
$G\setminus \{i\}$. By taking the union of the path from $k$ to
$q$ and the one from $q$ to $j$, one obtains then a directed path
from $k$ to $j$ in $G\setminus \{i\}$, as shown in Figure
\ref{fig:one_among_2_reverse}(c).}
\end{itemize}
In any of these three cases, there is thus a directed path from
$k$ or $l$ to $j$ in $G\setminus \{i\}$. As explained above, this
implies that one can perform the reverse atypical edge splitting
on $i$ if one cannot perform the reverse standard one.
\end{proof}

\subsection{Obtaining all minimally persistent graphs using $\A$}

Let $G=(V,E)$ be a minimally persistent and therefore minimally
rigid graph with more than two vertices. By Proposition
\ref{prop:Laman_min}, there holds $4\abs{V}-6 = \abs{E}=\sum_{i\in
V}d^-(i)+ d^+(i)$. Moreover, it can be shown (using Proposition
\ref{prop:Laman_min}) that such a graph never contains any vertex
with a total degree smaller than 2. A counting argument shows then
that it always contain at least one vertex with either $(d^-,d^+)
= (0,2)$, $(1,1)$ or $(1,2)$. In the first two cases, one can
perform on this vertex a reverse standard or atypical vertex
addition, while in the last case, it follows from Lemma
\ref{lem:one_among_2_reverse} that either a reverse standard edge
splitting or a reverse atypical one can be performed. It is thus
always possible to obtain a minimally persistent graph
$G'=(V',E')$ with $\abs{V'}=\abs{V}-1$ by performing an operation
of $\A^{-1}$ on $G$. Doing this recursively, one can obtains after
$\abs{V}-2$ operation a minimally persistent graph on two
vertices, i.e., a leader-follower seed. The reverse sequence of
operations allows then one to obtain $G$ from this seed. Since the
operations of $A$ and $A^{-1}$ preserve minimal persistence, all
the intermediate graph of such a sequence are persistent. We have
thus proved the following theorem:

\begin{thm}\label{thm:4op}
Every minimally persistent graph $G=(V,E)$ ($\abs{V}>1$) can be
obtained by performing $\abs{V}-2$ operations of $\A$ on an
initial leader-follower seed. Moreover, all the intermediate
graphs are minimally persistent.
\end{thm}

Using the same argument as for Corollary \ref{cor:tranfo_T_T^-1},
we obtain the following result.

\begin{cor}\label{cor:tranfo_A A^-1}
Every minimally persistent graph can be transformed into any other
minimally persistent graph using only operations of $\A \cup \A
^{-1}$.
\end{cor}
As in Section \ref{sec:obtention_with_T}, one can use the set $\A
\cup \A ^{-1}$ to define a distance on the minimally persistent
graphs (on more than one vertex). But, as a consequence of Theorem
\ref{thm:4op}, the distance between the graph $G=(V,E)$ and
$G'=(V',E')$ is never greater than $\abs{V}+ \abs{V'}
-4$.\\%

\begin{rem} The non-confined character of the atypical
edge splitting makes it more complicated to implement in an
autonomous agent formation. It can indeed involve the direction
reversal of a potentially large number of edges that are not
involved in the corresponding operation for undirected graph.
Proposition \ref{prop:nonconfined} states that there always is
such a non-confined operation in a set of generalized vertex
addition and edge splitting operations which allows one to build
all minimally persistent graphs. However, the example in Figure
\ref{fig:counterex_nonlocal} only requires one edge to be
reversed, and no example was found yet where it was necessary to
reverse more than one edge. There might thus exist a set of
operations having the same properties as $\A$ (with respect to the
building of all minimally persistent graphs) but in which the
non-confined operation only involves the reversal of a number of
edges bounded independently of $\abs{V}$.
\end{rem}

\begin{rem}
It is possible to show that among the four operations of $\A$
(resp. $\A^{-1}$), none can be removed without being replaced by
some alternative new operation if the operation set is to produce
all minimally persistent graphs (resp. contain for each minimally
persistent graph an operation that can be performed on it). For
each operation of $\A^{-1}$, one can indeed find a graph where
none of the three other operations can be performed. However, the
set of generalized vertex additions and edge splittings that we
present here is just one among the several sets that we have found
allowing one to build all minimally persistent graphs. It offers
the advantage that its non-confined operation has a more local
character than those contained in the other sets (which are not
described here).
\end{rem}

\subsection{Relations between $\A$ and $\T$}

We examine here the relation between the two sets $\A$ and $\T$.
Let $\X$ and $\Y$ be two sets of operations. We say that
$\X\leq\Y$ if all the operations of $\X$ can be implemented by a
sequence of operations of $\Y$. If $\X\leq\Y$ and $\X\geq\Y$, we
say that $\X=\Y$. If $\X\leq\Y$ and $\X \not=\Y$, we say that $\X<
\Y$. One can see
that $\X^{-1}\leq \Y^{-1}$ if and only if $\X\leq \Y$.\\%

\begin{lem}\label{lem:implem_ava}
An atypical vertex addition can be implemented using one standard
vertex addition and one edge reversal.
\end{lem}
\begin{proof}
Let $k$ and $j$ be two distinct vertices in a minimally persistent
graph such that $k$ has at least one degree of freedom. Instead of
adding a vertex $i$ and the edge $(i,j)$ and $(k,i)$ (atypical
vertex addition), one can equivalently add the vertex $i$ and the
edges $(i,j)$ and $(i,k)$ (standard vertex addition), and then
reverse the edge $(i,k)$ (edge reversal). Note that this edge
reversal can be performed because $k$ has a degree of freedom.
\end{proof}

\begin{lem}\label{lem:implem_aes}
An atypical edge splitting can be implemented using one standard
edge splitting and one or more edge reversal(s).
\end{lem}
\begin{proof}
Let $j$, $k$ and $l$ be three vertices of a minimally persistent
graph $G$ satisfying the conditions required to perform an
atypical edge splitting (see Section \ref{sec:description_4}), and
let $G'$ be the graph obtained by performing an atypical edge
splitting on $(k,l)$ in $G$ as represented in Figure
\ref{fig:repres_atypical operation}(b). By performing a standard
edge splitting on $(k,l)$ in $G$ such that the added vertex is
also connected to $j$, one obtains a minimally persistent graph
having the same underlying undirected graph as $G'$. It is then a
consequence of Proposition \ref{prop:class=undir_struc} that this
last graph can be obtained by a sequence of edge reversals.
\end{proof}

\begin{prop}\label{prop:A>T}
$\A<\T$, and equivalently $\A^{-1}< \T^{-1}$
\end{prop}
\begin{proof}
Observe first that all the operations of $\A$ increase the number
of vertices in the graph, while edge reversal does not. Thus edge
reversal cannot be implemented by a sequences of operations of
$\A$, and $\A \not \geq \T$.\\
Since the operations of $\S$ (standard vertex additions and edge
splitting) belong to both $\A$ and $\T$, and since by Lemmas
\ref{lem:implem_ava} and \ref{lem:implem_aes}, the operations of
$\A \setminus \S$ (atypical vertex addition and atypical edge
splitting) can be implemented using operations of $\T$, we have
$\A \leq \T$ , which together with $A \not \geq T$ implies that
$\A<\T$.
\end{proof}

Since the set $\T$ of operations is more powerful than $\A$,
Theorem \ref{thm:4op} is a stronger result than Theorem
\ref{thm:3op}. However, if we look at the sets containing both
normal and inverse operations, the results are different. Suppose
indeed that a graph $G'$ is obtained by performing an operation of
$\T\cup \T^{-1}$ on a minimally persistent graph $G$. Since both
graphs are minimally persistent, Corollary \ref{cor:tranfo_A A^-1}
implies that $G'$ can also be obtained by applying a sequence of
operations of $\A\cup\A^{-1}$ on $G$. Any operation of $\T\cup
\T^{-1}$ can thus be implemented by a sequence of operations of
$\A\cup\A^{-1}$. Conversely, any operation of $\A\cup\A^{-1}$ can
be implemented by a sequence of operations of $\T\cup \T^{-1}$. We
have thus shown the following result:

\begin{prop}
$\A\cup \A^{-1} = \T\cup \T^{-1}$
\end{prop}

Both sets $\A$ and $\T$ allow one to enumerate exhaustively all
minimally persistent graphs. However, as explained in Section
\ref{sec:min_rig_graph}, since the sequence that can build a
certain minimally persistent graph is not unique, this enumeration
can allow one to compute an upper bound on the number of minimally
persistent graphs having a certain number of vertices, but not
their exact number.

\section{Conclusions and future work} \label{sec:conclusions}

In this paper, we have extended the Henneberg sequence concept to
directed graphs. From an autonomous agent point of view, this
provides a systematic approach to sequentially obtain or
reorganize a minimally persistent agent formation. We also exposed
some natural restrictions to these extensions, the main one being
the impossibility of building all minimally persistent graphs
using only confined generalized vertex additions or edge
splittings.\\%

We proposed two sets of operations, each of which allows one to
obtain any minimally persistent graph from a leader-follower seed.
The first one ($\T$ in Section \ref{sec:third_op}) contains the
two standard vertex additions and edge splittings already
introduced in \cite{ErenWhiteleyAndersonMorseBelhumeur:2005} and a
purely directed operation (i.e. a neutral operation from an
undirected point of view). The second set ($\A$ in Section
\ref{sec:four_non_local}) contains, in addition to the two
standard operations, two atypical versions of them, among which
one is not a confined operation. It involves indeed the reversal
of a path of undetermined length in the graph. However, it is
still an open question to know if similar results could be
obtained using operations involving a number of reversals fixed or
bounded independently of the size of the graph. Note that for the
second set, the number of operations required to build a minimally
persistent graphs is uniquely determined by the size
of the graph.\\%

From an autonomous agent point of view, it would be useful to
study how these various operations could be performed in a
decentralized way in order to efficiently construct or reorganize
a formation. For this purpose, the operations of $\T$ could be
preferred for their simplicity and because the successive
modifications are only local modifications. This study could also
imply the development of an optimal algorithm (using one of the
two sets of operations) to reorganize a persistent formation. An
improvement could come from the use of an even more simple
operation such as edge reorientation, which would consist in
changing the arrival point of an edge. However, the conditions
under which minimal rigidity is preserved by this operation are
not known yet.\\%

As a final remark, note that we have only focused on the
transformations of minimally persistent graph into other minimally
persistent graphs. Several practical issues concerning autonomous
agent formations arise relating to the merging of such graphs, or
their repair after the loss of some vertices or edges.  It would
thus be worthwhile to study these particular problems as well.

\bibliography{references}

\begin{thebibliography}{10}

\bibitem{BaillieulSuri:2003}
J.~Baillieul and A.~Suri.
\newblock Information patterns and hedging brockett's theorem in controlling
  vehicle formations.
\newblock In {\em Proc. of the 42nd IEEE Conf. on Decision and Control},
  volume~1, pages 556--563, Hawaii, December 2003.

\bibitem{DasSpletzerKumarTaylor:2002}
A.~Das, J.~Spletzer, V.~Kumar, and C.~Taylor.
\newblock Ad hoc networks for localization and control.
\newblock In {\em Proc. of the 41st IEEE Conf. on Decision and Control},
  volume~3, pages 2978--2983, Las Vegas, NV, December 2002.

\bibitem{ErenWhiteleyAndersonMorseBelhumeur:2005}
T.~Eren, B.D.O. Anderson, A.S. Morse, and P.N. Belhumeur.
\newblock Information structures to secure control of rigid formations with
  leader-follower structure.
\newblock In {\em Proc. of the American Control Conference}, pages 2966--2971,
  Portland, Oregon, June 2005.

\bibitem{HendrickxAndersonBlondel:2005}
J.M. Hendrickx, B.D.O. Anderson, and V.D. Blondel.
\newblock Rigidity and persistence of directed graphs.
\newblock In {\em Proceedings of the 44th IEEE Conference on Decision and
  Control}, Seville, Spain, December 2005.

\bibitem{HendrickxAndersonDelvenneBlondel:2005}
J.M. Hendrickx, B.D.O. Anderson, J.-C. Delvenne, and V.D. Blondel.
\newblock Directed graphs for the analysis of rigidity and persistence in
  autonomous agents systems.
\newblock {\em To appear in International Journal of Robust and Nonlinear
  Control's special issue on Communicating-Agent Systems}.

\bibitem{HendrickxFidanYuAndersonBlondel:2005_conf}
J.M. Hendrickx, B.~Fidan, C.~Yu, B.D.O. Anderson, and V.D. Blondel.
\newblock Rigidity and persistence of three and higher dimensional formations.
\newblock In {\em Proceedings of the First International Workshop on
  Multi-Agent Robotic Systems (MARS 2005)}, pages 39--46, Barcelona, Spain,
  September 2005.

\bibitem{Henneberg:11}
L.~Henneberg.
\newblock Die graphische {S}tatik der starren {S}ysteme.
\newblock Leipzig, 1911.

\bibitem{Laman:70}
G.~Laman.
\newblock On graphs and rigidity of plane skeletal structures.
\newblock {\em J. Engrg. Math.}, 4:331--340, 1970.

\bibitem{OlfatiMurray:2002}
R.~Olfati-Saber and R.M Murray.
\newblock Graph rigidity and distributed formation stabilization of
  multi-vehicle systems.
\newblock In {\em Proceedings of the 41st Conference on Decision and Control},
  volume~3, pages 2965--2971, Las Vegas, NV, December 2002.

\bibitem{TayWhiteley:85}
T.~Tay and W.~Whiteley.
\newblock Generating isostatic frameworks.
\newblock {\em Structural Topology}, (11):21--69, 1985.

\bibitem{Whiteley:96}
W.~Whiteley.
\newblock Some matroids from discrete applied geometry.
\newblock In {\em Matroid theory (Seattle, WA, 1995)}, volume 197 of {\em
  Contemp. Math.}, pages 171--311. Amer. Math. Soc., Providence, RI, 1996.

\bibitem{YuHendrickxFidanAnderson:2005}
C.~Yu, J.M. Hendrickx, B.~Fidan, and B.D.O. Anderson.
\newblock Structural persistence of three dimensional autonomous formations.
\newblock In {\em Proceedings of the First International Workshop on
  Multi-Agent Robotic Systems (MARS 2005)}, pages 47--55, Barcelona, Spain,
  September 2005.

\bibitem{YuHendrickxFidanAndersonBlondel:2005}
C.~Yu, J.M. Hendrickx, B.~Fidan, B.D.O. Anderson, and V.D. Blondel.
\newblock Three and higher dimensional autonomous formations: Rigidity,
  persistence and structural persistence.
\newblock {\em To appear in Automatica}.

\end{thebibliography}

\end{document}